\documentclass[12pt]{article}

\usepackage{graphicx,psfrag,mathrsfs,bbm,amssymb,here,amsmath,cite}

\newcommand{\be}{\begin{equation}}
\newcommand{\ee}{\end{equation}}
\newcommand{\capslash}[1]{{#1}\mskip -12mu / \mskip +4 mu}

\newcommand{\msp}{\;\;\;\;\;}
\newcommand{\boldm}[1]{\mbox{\boldmath ${#1}$}}
\newcommand{\lf}{\left}
\newcommand{\rg}{\right}

\graphicspath{{figures/}}
\DeclareGraphicsExtensions{.eps.gz,.eps,.ps,.ps.gz}
              
\begin{document}


\begin{flushright}
  LC-TH-2005-007 \\
  HD-THEP-05-18 \\
  hep-ph/0508132 
\end{flushright}

\vspace{\baselineskip}

\begin{center}
\textbf{\LARGE Effective-Lagrangian approach \\[0.3em]
        to \boldm{\gamma \gamma \rightarrow WW};\\[0.3em]
I: Couplings and amplitudes \\}
\vspace{4\baselineskip}
{\sc O.~Nachtmann\footnote{email:
O.Nachtmann@thphys.uni-heidelberg.de}, F.~Nagel\footnote{email:
F.Nagel@thphys.uni-heidelberg.de}, M.~Pospischil\footnote{Now at
CNRS UPR 2191, 1 Avenue de la Terrasse, F-91198 Gif-sur-Yvette,
France,\\\hspace*{.64cm}email: Martin.Pospischil@iaf.cnrs-gif.fr} and 
A.~Utermann\footnote{email:
A.Utermann@thphys.uni-heidelberg.de}
}\\
\vspace{1\baselineskip}
\textit{Institut f\"ur Theoretische Physik, Philosophenweg 16, D-69120
Heidelberg, Germany}\\
\vspace{2\baselineskip}
\textbf{Abstract}\\
\vspace{1\baselineskip}
\parbox{0.9\textwidth}{We consider the gauge-boson sector of a locally
$SU(2)\times U(1)$ invariant effective Lagrangian with ten
dimension-six operators added to the Lagrangian of the Standard Model.
These operators induce anomalous three- and four-gauge-boson couplings
and an anomalous
\mbox{$\gamma \gamma H$}~coupling.  In the framework of this effective
Lagrangian we calculate the helicity amplitudes and differential and
total cross sections for the process \mbox{$\gamma \gamma \rightarrow
WW$} at a photon collider.  We give relations between different parts
of the amplitudes that show which linear combinations of anomalous
couplings are measurable in this reaction. The transformation
properties of the differential cross section under~$CP$ are
discussed. We find that three linear combinations of $CP$~conserving
and of $CP$~violating couplings can be measured independently of the
photon polarisation in~\mbox{$\gamma \gamma \rightarrow WW$}.}
\end{center}
\vspace{\baselineskip}

\pagebreak

\tableofcontents

\pagebreak

\section{Introduction}
\label{sec-intro}

Given the large variety of particle-physics models that claim to
replace the Standard Model~(SM) at a high energy scale~$\Lambda$, it
is a vital approach for precision experiments at a future
\mbox{$e^+e^-$} linear collider~ILC with a design like
TESLA~\cite{Richard:2001qm} or CLIC~\cite{Ellis:1998wx} to probe
new-physics effects in a model-independent way. A $\gamma\gamma$
collider---where two high-energy photons are obtained through Compton
backscattering off high-energy electrons--- extends the physics
potential of an ILC substantially. Such a photon-collider option
is planned for example at \mbox{$e^+e^-$}~machines like
TESLA~\cite{Badelek:2001xb} or CLIC~\cite{Burkhardt:2002vh}. The~SM is
very successful at energies up to the electroweak scale, set by the
vacuum expectation value,
\mbox{$v \approx 246$}~GeV, of the SM-Higgs-boson field.  Assuming
\begin{equation}
\label{eq-scales}
\Lambda \gg v
\end{equation}
the new-physics effects can be taken into account using an effective
Lagrangian that consists of the SM~Lagrangian supplemented by
operators of dimension higher than four, see
\cite{Buchmuller:1985jz,Leung:1984ni} and references therein.  We
adopt the approach of~\cite{Buchmuller:1985jz}, where all operators up
to dimension six are constructed that contain only SM fields and are
invariant under the SM gauge symmetry
\mbox{$SU(3) \times SU(2) \times U(1)$}.  In a preceding
work~\cite{Nachtmann:2004ug} we have extensively discussed the gauge-boson
sector of such a Lagrangian,
\begin{equation}
\label{eq-Leff}
\mathscr{L}_{\rm eff} = \mathscr{L}_{0} + \mathscr{L}_{2}\,,
\end{equation}
where $\mathscr{L}_{0}$ is the Lagrangian of the~SM (for our
conventions see App.~\ref{sec-conv}) and the second term
\begin{equation}
\begin{split}
\label{eq-dim6op}
\mathscr{L}_{2} & = 
\Big(h_W O_W +
h_{\tilde{W}} O_{\tilde{W}} +
h_{\varphi W} O_{\varphi W} +
h_{\varphi \tilde{W}} O_{\varphi \tilde{W}} +
h_{\varphi B} O_{\varphi B} +
h_{\varphi \tilde{B}} O_{\varphi \tilde{B}} \\
 &  \qquad + 
h_{W\! B} O_{W\! B} +
h_{\tilde{W}\! B} O_{\tilde{W}\! B} +
h_{\varphi}^{(1)} O_{\varphi}^{(1)} +
h_{\varphi}^{(3)} O_{\varphi}^{(3)}\Big)/v^2\,,
\end{split}
\end{equation}
contains all dimension-six operators that either consist only of electroweak
	gauge-boson fields or that contain both gauge-boson fields and the SM-Higgs
field.  There is no gauge invariant dimension-five operator that can be
constructed out of these fields.  We list the definitions of the operators
of~$\mathscr{L}_{2}$ in App.~\ref{sec-conv}.  Four operators are
$CP$~violating, namely those containing the dual $W$- or $B$-field strengths;
they are denoted by a tilde on the subscripts in~(\ref{eq-dim6op}).  We have
divided by~$v^2$ in order to render the coupling constants~$h_i$
dimensionless.  The~$h_i$ are subsequently called anomalous couplings.  We have
\be 
\label{eq-orderh}
h_i \sim O\lf(v^2/\Lambda^2\rg)\,.
\ee
The new operators~(\ref{eq-dim6op}) have an impact on a large number of
electroweak precision observables as calculated
in~\cite{Nachtmann:2004ug,Leung:1984ni,Hagiwara:1992eh}.
It is particularly interesting to study the rich phenomenology induced
by the Lagrangian~(\ref{eq-Leff}) at an ILC both in the
high-energy \mbox{$e^+ e^-$} and~\mbox{$\gamma \gamma$} modes, and in
the Giga-$Z$ mode.

There are other ways to parametrise new-physics effects. For
instance in a form-factor approach for the \mbox{$\gamma WW$} and
\mbox{$ZWW$} couplings anomalous effects are para\-metrised by 28 real
parameters if one also allows for imaginary parts in the form
factors~\cite{Hagiwara:1986vm}. In such a framework no anomalous
contributions at the fermion-boson vertices or at the boson
propagators occur. The sensitivity to the anomalous triple-gauge-boson
couplings in the reaction \mbox{$e^+ e^- \rightarrow WW$} at an
ILC for different beam polarisations has been studied in detail
in~\cite{Diehl:2002nj}. There are ways to translate the bounds on the
anomalous couplings from the form-factor approach to the
effective-Lagrangian approach~\cite{Nachtmann:2004ug}.  In the same
reference also the advantages of the different approaches are surveyed
and a discussion of the relation to other work on anomalous
electroweak gauge-boson couplings in $e^+e^-$ annihilation
\cite{Hagiwara:1986vm,Richard:2001qm,Diehl:1993br,Diehl:2002nj,Kuss:1997mf,Gaemers:1978hg,Berends:1997av,Menges:2001gg,Abe:2001wn,Bozovic-Jelisavcic:2002ta} is given.

In the form-factor approach implications from a future $\gamma\gamma$
collider for anomalous couplings were discussed in
\cite{Baillargeon:1997rz,Bozovic-Jelisavcic:2002ta,Tupper:1980bw,Belanger:1992qi,Marfin:2003jg,Belanger:hp,Bredenstein:2004ef,Bredenstein:2005zk,Banin:1998ap}. At tree level anomalous triple gauge-boson couplings
from the $\gamma WW$ vertex but not from the $ZWW$ vertex contribute. In
\cite{Bozovic-Jelisavcic:2002ta} it was argued that
the precision for the measurement of the $\gamma WW$ coupling is
comparable in the $e^+e^-$ and the $\gamma\gamma$ modes. 
The $\gamma\gamma WW$ vertex contributes already at tree level at a
photon collider. Hence also implications from anomalous quartic
gauge-boson couplings can be investigated. Certain analyses of
anomalous couplings at a $\gamma\gamma$ collider are focused on
anomalous triple gauge-boson couplings
\cite{Bozovic-Jelisavcic:2002ta,Tupper:1980bw}, on anomalous quartic
gauge-boson couplings \cite{Belanger:1992qi,Marfin:2003jg} and on
$CP$-violating gauge-boson couplings \cite{Belanger:hp,Choi:1996xt}. In
\cite{Baillargeon:1997rz,Bredenstein:2004ef,Bredenstein:2005zk} the finite width of the
$W$ boson and virtual and real corrections \cite{Bredenstein:2005zk}
are also taken into account. For a discussion of the anomalous
$\gamma\gamma H$ vertex in the $\gamma\gamma$ and $e\gamma$ modes at
an ILC see \cite{Banin:1998ap}. Implications for the process
$\gamma\gamma\to WW$ induced by some of the dimension-six operators in
(\ref{eq-dim6op}) were investigated in
\cite{Gounaris:1995mc,Choi:1996xt}.  

In our present work we study, in the framework of the
Lagrangian~(\ref{eq-Leff}), the process \mbox{$\gamma\gamma
\rightarrow WW \rightarrow 4$}~fermions. In the reaction
\mbox{$\gamma\gamma\rightarrow WW$} anomalous contributions to the
\mbox{$\gamma WW$}, \mbox{$\gamma \gamma WW$} and \mbox{$\gamma \gamma
H$} vertices can be studied.  We do this by applying the Lagrangian
(\ref{eq-Leff}) and (\ref{eq-dim6op}) as calculated
in~\cite{Nachtmann:2004ug} in terms of the physical fields~$A$, $Z$,
$W^{\pm}$ and~$H$. Finally we expand to linear order in the anomalous
couplings. Throughout this and the companion paper II
\cite{Nachtmann:opti}, we use the parameter scheme~$P_W$ (see
Sect. 4.2 of \cite{Nachtmann:2004ug}), which contains the $W$~boson
mass $m_W$ as an input parameter. In Sect. 4.1 of
\cite{Nachtmann:2004ug} we discussed the scheme $P_Z$, which contains the $Z$~boson mass
$m_Z$ as an input parameter. There the $h_i$ would modify the $W$~mass
and therefore the kinematics of the reaction, which is highly
inconvenient.  Let us mention that in our approach, where we use the
effective Lagrangian (\ref{eq-Leff}) to construct all required
interactions, no further assumptions are needed to relate the cross
sections for the $e^+e^-$ and the $\gamma\gamma$ modes. On the other
hand in the form-factor approach one has to add for every contributing
vertex new form factors. Hence it is getting more involved to
calculate combined constraints from the $e^+e^-$ and the $\gamma\gamma$
modes on anomalous couplings. In the companion paper
\cite{Nachtmann:opti} we will compare the sensitivity on the couplings
in (\ref{eq-dim6op}) from the $e^+e^-$ and $\gamma\gamma$ modes and
high precision observables. To this end we shall calculate in
\cite{Nachtmann:opti} the minimum errors on the anomalous couplings
obtainable in \mbox{$\gamma \gamma\rightarrow WW$} by means of optimal
observables~\cite{Atwood:1991ka,Diehl:1993br}.

Our paper is organised as follows: In Sect.~\ref{sec-cross} we derive
the helicity amplitudes and the differential cross section for the
process \mbox{$\gamma \gamma \rightarrow WW$} with fixed
c.m.~energy. In Sect.~\ref{sec-dissymm} the discrete-symmetry
properties of the anomalous interactions are explained. We present our
conclusions in Sect.~\ref{sec-conc}. In App.~\ref{sec-conv} we give
our conventions for the SM~Lagrangian and list the additional
dimension-six operators of the effective Lagrangian. Those Feynman
rules derived from the effective Lagrangian that are needed for the
calculation of the process
\mbox{$\gamma \gamma \rightarrow WW \rightarrow 4$}~fermions are listed in
App.~\ref{sec-feynm}. In App.~\ref{sec-pola} we give our conventions for
particle momenta and polarisation vectors.  The analytic results for the
helicity amplitudes of the production process \mbox{$\gamma \gamma \rightarrow
WW$} are given in App.~\ref{sec-helggww}.


\section{Cross section for \boldm{\gamma \gamma \rightarrow WW
\rightarrow} 4 fermions}
\label{sec-cross}
\setcounter{equation}{0}

\begin{figure}[h]
\centering
\includegraphics[totalheight=5cm]{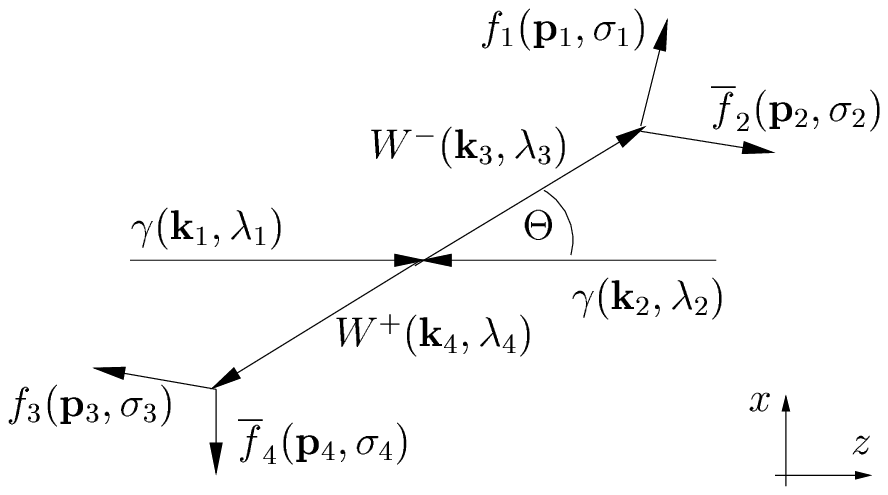}
\caption{\label{fig:conv}Conventions for particle momenta and helicities.}
\end{figure}

In this section we derive the spin-averaged differential cross section
for the process
\be
\label{eq-process}
\gamma \gamma \rightarrow W^- W^+ \rightarrow (f_1 \overline{f}_2)(f_3 
\overline{f}_4)
\ee
in the scheme~$P_W$.  The final-state fermions in~(\ref{eq-process})
are leptons or quarks.  We consider this reaction for fixed photon
energies in the framework of the effective Lagrangian~(\ref{eq-Leff}).
The case where the initial photons are not monochromatic but have
Compton energy-spectra raises additional complications with the
kinematical reconstruction of the final state.  This will be
considered in~\cite{Nachtmann:opti} applying the methods presented
in~\cite{Nachtmann:2004fy}.  Our notation for particle momenta and
helicities is shown in Fig.~\ref{fig:conv}.
The production of the $W$~bosons is described in the \mbox{$\gamma \gamma$
c.m.\ frame}.  Our coordinate axes are chosen such that the \mbox{$WW$}-boson
production takes place in the $x$-$z$ plane, the photon momentum ${\bf k}_1$
points in the positive $z$-direction and the $y$ unit vector is given by
\mbox{$\boldm{\hat{e}}_y = ({\bf k}_1 \times {\bf k}_3)/|{\bf k}_1 \times {\bf
k}_3|$}.  In the \mbox{$\gamma \gamma$}~c.m.~frame and at a given c.m.~energy
$\sqrt{s}$, a pure initial state of two photons is uniquely specified by the
photon helicities:
\be
|\lambda_1 \lambda_2 \rangle = |\gamma ({\bf k}_1, \lambda_1) \gamma
({\bf k}_2, \lambda_2) \rangle\;\;\;\;\;\;\;\;\left(\lambda_1,
\lambda_2 = \pm 1\right)\,.
\ee
Then the differential cross section for an unpolarised initial state is
\be
\label{eq-dsigma}
{\rm d}\sigma = \frac{1}{2s}\; \frac{1}{4} \sum_{\lambda_1, \lambda_2}
|\langle f | \mathcal{T} | \lambda_1 \lambda_2 \rangle |^2 \;{\rm d}\Gamma\,,
\ee
where $\mathcal{T}$ is the transition operator, $| f \rangle = |f_1 ({\bf
p}_1, \sigma_1)\, \overline{f}_2 ({\bf p}_2, \sigma_2)\, f_3 ({\bf p}_3,
\sigma_3)\, \overline{f}_4 ({\bf p}_4, \sigma_4) \rangle$ is the final state
and the phase-space measure for final states is as usual
\be
\label{eq-phase}
{\rm d}\Gamma = \left( \prod^4_{i=1} \frac{{\rm d}^3 p_i}{(2 \pi)^3 2
p^0_i} \right) (2 \pi)^4 \delta^{(4)} \left(k_1+k_2-\sum_{i=1}^4 p_i\right)\,.
\ee
Using the narrow-width approximation for the $W$~bosons and considering all
final-state fermions to be massless, we obtain
\be
\label{eq-diffs}
\frac{{\rm d}\sigma}{{\rm d}\!\cos \Theta \; {\rm d}\!\cos \vartheta \;  {\rm
d}\varphi \; {\rm d}\!\cos \overline{\vartheta} \;  {\rm d}\overline{\varphi}}
= \frac{3\beta}{2^{13} \pi^3 s} \, B_{12} \, B_{34} \, \mathcal{P}^{\lambda_3
\lambda_4}_{\lambda^\prime_3 \lambda^\prime_4} \, \mathcal{D}^{\lambda_3}_{\lambda^\prime_3} \,
\overline{\mathcal{D}}^{\lambda_4}_{\lambda^\prime_4}\;, 
\ee
where summation over repeated indices is implied and \mbox{$\beta = (1-4 m_W^2
/ s)^{1/2}$} is the velocity of each $W$~boson in the \mbox{$\gamma
\gamma$}~c.m.~frame.  The branching ratio for the decay
\mbox{$W \rightarrow f_i \overline{f}_j$} is denoted by $B_{ij}$. The $W$~helicity states are
defined in the coordinate system shown in Fig.~\ref{fig:conv}. For the
definition of the polarisation vectors see App.~\ref{sec-pola}. The
polar angle between the positive $z$-axis and the $W^-\!$~momentum is
denoted by~$\Theta$.  The cross section does not depend on the
azimuthal angle of the $W^-\!$~momentum due to rotational invariance.
The respective frames for the decay tensors are defined by a rotation
by $\Theta$ about the $y$-axis of the frame in Fig.~\ref{fig:conv}
such that the $W^-\!$ ($W^+$) momentum points in the new positive
(negative) $z$-direction and a subsequent rotation-free boost into
the c.m.~system of the corresponding $W$~boson.  The spherical
coordinates $\vartheta, \varphi$ and
$\overline{\vartheta},\overline{\varphi}$ are those of the $f_1$- and
$\overline{f}_4$-momentum directions, respectively.  The production
and decay tensors in (\ref{eq-diffs}) are given by
\begin{align}
\label{eq-prodpro}
\mathcal{P}^{\lambda_3 \lambda_4}_{\lambda^\prime_3 \lambda^\prime_4} (\Theta) & = 
\sum_{\lambda_1, \lambda_2} \mathcal{M}(\lambda_1, \lambda_2; \lambda_3,
\lambda_4) \mathcal{M}^*(\lambda_1, \lambda_2; \lambda^\prime_3,\lambda^\prime_4)\,, \\
\mathcal{M}(\lambda_1,\lambda_2;\lambda_3,\lambda_4)&\equiv
\langle W^-({\bf k}_3, \lambda_3)\,W^+({\bf k}_4,\lambda_4)|\mathcal{T}|\gamma({\bf k_1},\lambda_1)\,\gamma({\bf k_2},\lambda_2)\rangle\,,\\
\label{eq-decay12}
\mathcal{D}^{\lambda_3}_{\lambda^\prime_3}(\vartheta, \varphi) & = 
l^{\phantom{*}}_{\lambda_3}
l^*_{\lambda^\prime_3}\;,\\
\label{eq-decay34}
\overline{\mathcal{D}}^{\lambda_4}_{\lambda^\prime_4}(
\overline{\vartheta},\overline{\varphi}) & = 
\overline{l}^{\phantom{*}}_{\lambda_4} \overline{l}^{\,*}_{\lambda^\prime_4}\;,
\end{align}
where we have suppressed the phase-space variables on the right hand side.
The functions occurring in the decay tensors are listed in
App.~\ref{sec-helggww}.

\begin{table}
\centering
\begin{tabular}{rccccccccccc}
\hline
 &&&&&&&&&&&\\[-.45cm]
 & SM & $h_W$ & $h_{\tilde{W}}$ & $h_{\varphi W}$ & $h_{\varphi \tilde{W}}$
 & $h_{\varphi B}$ & $h_{\varphi \tilde{B}}$ & $h_{W\! B}$ &
 $h_{\tilde{W}\! B}$ & $h_{\varphi}^{(1)}$ &
 $h_{\varphi}^{(3)}$\\[.1cm]
\hline
 &&&&&&&&&&&\\[-.43cm]
$\gamma WW$ & $\surd$ & $\surd$ & $\surd$ & & & & &
 $\surd$ & $\surd$ & &\\
$ZWW$ & $\surd$ & $\surd$ & $\surd$ & & & & & $\surd$ & $\surd$ &  & $P_Z$\\
$\gamma \gamma WW$ & $\surd$ & $\surd$ & $\surd$ & & & & & & & &\\
$\gamma \gamma H$ & & & & $\surd$ & $\surd$ & $\surd$ & $\surd$ & $\surd$ &
 $\surd$ & &\\[.05cm]
\hline
\end{tabular}
\caption{\label{tab:vertices}Contributions of the SM Lagrangian and of
the anomalous operators~(\ref{eq-dim6op}) to different vertices in
order~\mbox{$O(h)$}.  The coupling~$h_{\varphi}^{(3)}$ contributes to the
$ZWW$~vertex in the scheme~$P_Z$ but not in~$P_W$.}
\end{table}

\begin{figure}
\centering
\includegraphics[totalheight=2.9cm]{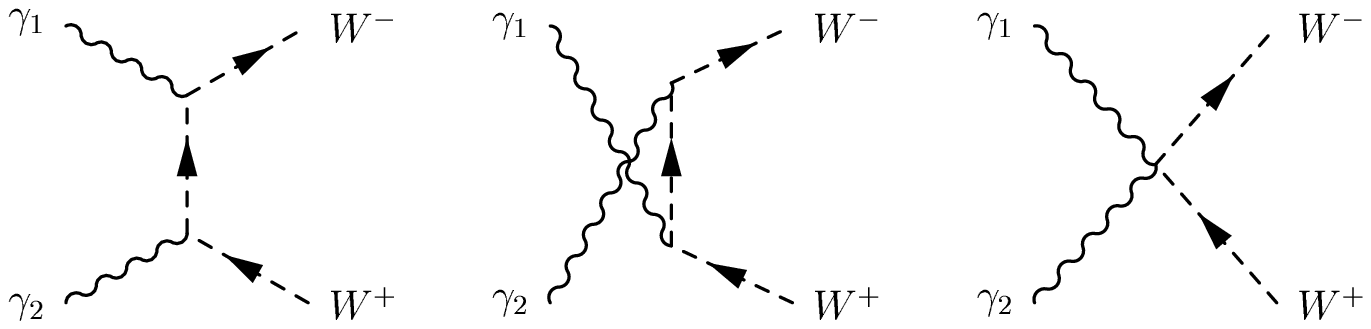}
\caption{\label{fig:feyndiag}SM diagrams for \mbox{$\gamma \gamma
\rightarrow WW$}.}  
\vspace{1.5cm}
\includegraphics[totalheight=5.8cm]{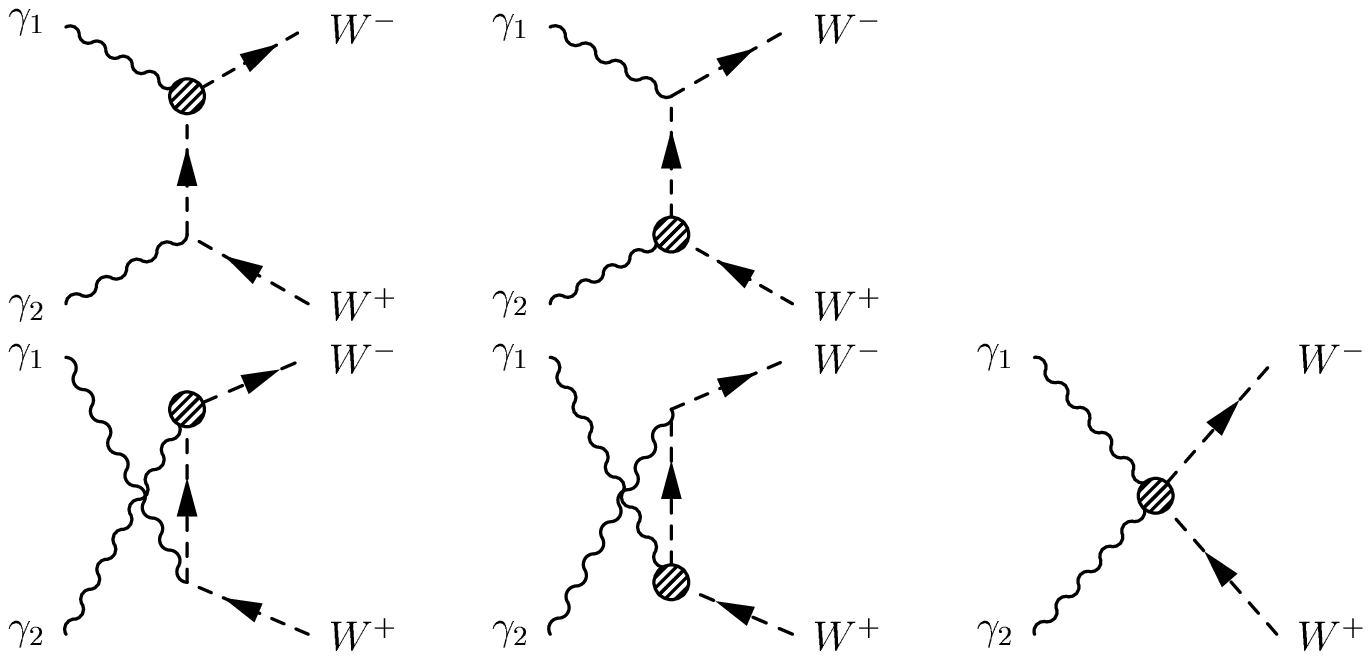}
\caption{\label{fig:anfeyndiag}Diagrams with anomalous triple- or
  quartic-gauge-boson couplings for \mbox{$\gamma \gamma \rightarrow WW$}.}
\vspace{1.5cm} 
\includegraphics[totalheight=2.9cm]{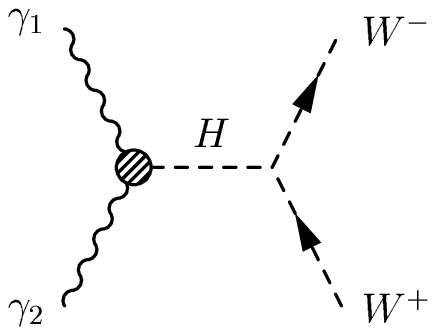}
\caption{\label{fig:anHdiag}Diagram with anomalous 
$\gamma \gamma H$ coupling for \mbox{$\gamma \gamma \rightarrow WW$}.}
\end{figure}

To first order in the anomalous couplings the production amplitude
$\mathcal{M}$ contains the SM diagrams (Fig.~\ref{fig:feyndiag}), diagrams
containing one anomalous triple- or quartic-gauge-boson vertex
(Fig.~\ref{fig:anfeyndiag}), and an $s$-channel Higgs-boson exchange
(Fig.~\ref{fig:anHdiag}).  The Feynman rules that are necessary to compute
these diagrams are listed in App.~\ref{sec-feynm}.  Tab.~\ref{tab:vertices}
shows which operators contribute to the three kinds of anomalous vertices in
Figs.~\ref{fig:anfeyndiag} and~\ref{fig:anHdiag}.
We expand the production amplitudes to first order in the anomalous
couplings:
\be
\label{eq-Mdecomp}
\mathcal{M} = \mathcal{M}_{\rm SM} 
+ \sum_i h_i \mathcal{M}_i \; + \; O(h^2)\,,
\ee
where all particle momenta and helicities are suppressed.  $\mathcal{M}_{\rm
SM}$ is the SM tree-level amplitude and \mbox{$i = W, \tilde{W}, \varphi W,
\varphi \tilde{W}, \varphi B, \varphi \tilde{B}, W\! B, \tilde{W}
\mskip -3mu B$}.  The couplings~$h_{\varphi}^{(1)}$ and~$h_{\varphi}^{(3)}$ do
not enter the amplitudes~(\ref{eq-Mdecomp}) to first order.  The different
terms on the right hand side of~(\ref{eq-Mdecomp}) for the various helicity
combinations of the incoming and outgoing gauge bosons are listed in
App.~\ref{sec-helggww}.  We find relations between amplitudes
corresponding to different anomalous couplings. These relations
depend on combinations of input parameters of $P_W$,
\begin{equation}
 s_1^2\equiv\frac{e^2}{4\,\sqrt{2}\,G_F\,m_W^2}\,,\qquad 
c_1^2\equiv 1-s_1^2\,, \label{s1c1}
\end{equation}
see section 4.2 of \cite{Nachtmann:2004ug}. Up to corrections from
anomalous couplings $s_1$ is the sine of the weak mixing angle. Two of
these relations are independent of the photon or $W$~helicities:
\begin{align}
 s_1^2\,\mathcal{M}_{\varphi B} & 
=c_1^2\,\mathcal{M}_{\varphi W}\,,
\label{Mphirel}
\\
\label{Mphitilderel}
 s_1^2\,\mathcal{M}_{\varphi \tilde{B}} & 
=c_1^2\,\mathcal{M}_{\varphi \tilde{W}}\,,
\end{align}
where all helicities and momenta are understood to be equal on both sides.
Hence the corresponding four anomalous couplings do not appear in the
amplitudes in an independent way but only as linear combinations
\begin{align}
 h_{\varphi W\! B}
&\equiv s_1^2\,h_{\varphi W} 
      + c_1^2\,h_{\varphi B}\,,
\label{eq-hphiWB}
\\
\label{eq-hphiWtildeB}
h_{\varphi \tilde{W}\!\tilde{B}} 
&\equiv s_1^2\,h_{\varphi \tilde{W}}
       +c_1^2\,h_{\varphi \tilde{B}}\,.
\end{align}
Another relation depends on the photon helicity~$\lambda_1$:
\be
\mathcal{M}_{\varphi \tilde{W}} = 2 i \lambda_1 \,\mathcal{M}_{\varphi W}\,.
\ee
Notice that the anomalous couplings do not modify the couplings of the
$W$~boson to fermions in the scheme~$P_W$, see~\cite{Nachtmann:2004ug}. Thus
there are no modifications of the decay amplitudes of the $W$~bosons due to
the~$h_i$.  We conclude that the differential cross section of \mbox{$\gamma
\gamma \rightarrow WW$} is sensitive to the anomalous couplings $h_W$,
$h_{\tilde{W}}$, $h_{\varphi W\! B}$, $h_{\varphi
\tilde{W}\!\tilde{B}}$, $h_{W\! B}$ and~$h_{\tilde{W}\! B}$. The
couplings $h_\varphi^{(1)}, h_\varphi^{(3)}$ and the orthogonal
combinations to (\ref{eq-hphiWB}) and (\ref{eq-hphiWtildeB}), that is
\begin{align}
  h^\prime_{\varphi W\! B} & 
=  c_1^2 h_{\varphi W}-s_1^2 h_{\varphi B}\,,
\label{eq-hphiWBprime}
\\
\label{eq-hphiWtildeBprime}
h^\prime_{\varphi \tilde{W}\!\tilde{B}} & 
= c_1^2 h_{\varphi \tilde{W}} -s_1^2 h_{\varphi\tilde{B}} \,,
\end{align}
do not enter in the expressions for the amplitudes of 
$\gamma\gamma\to WW$
 due to (\ref{Mphirel}) and (\ref{Mphitilderel}). Thus three $CP$
conserving couplings and one $CP$ violating coupling are
unmeasurable in $\gamma\gamma\to WW$.

\begin{figure}
\centering
\includegraphics[totalheight=9cm]{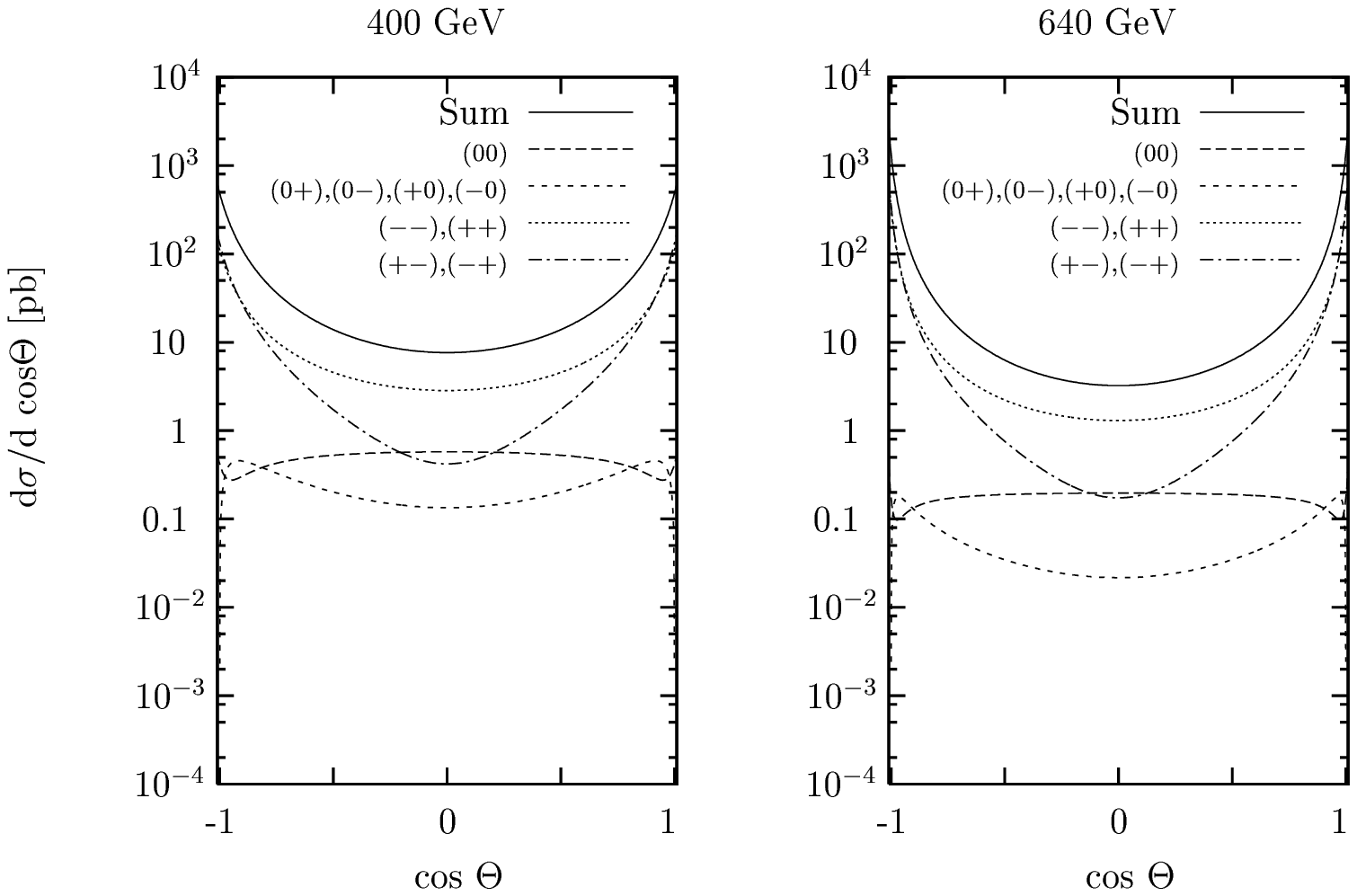}
\includegraphics[totalheight=9cm]{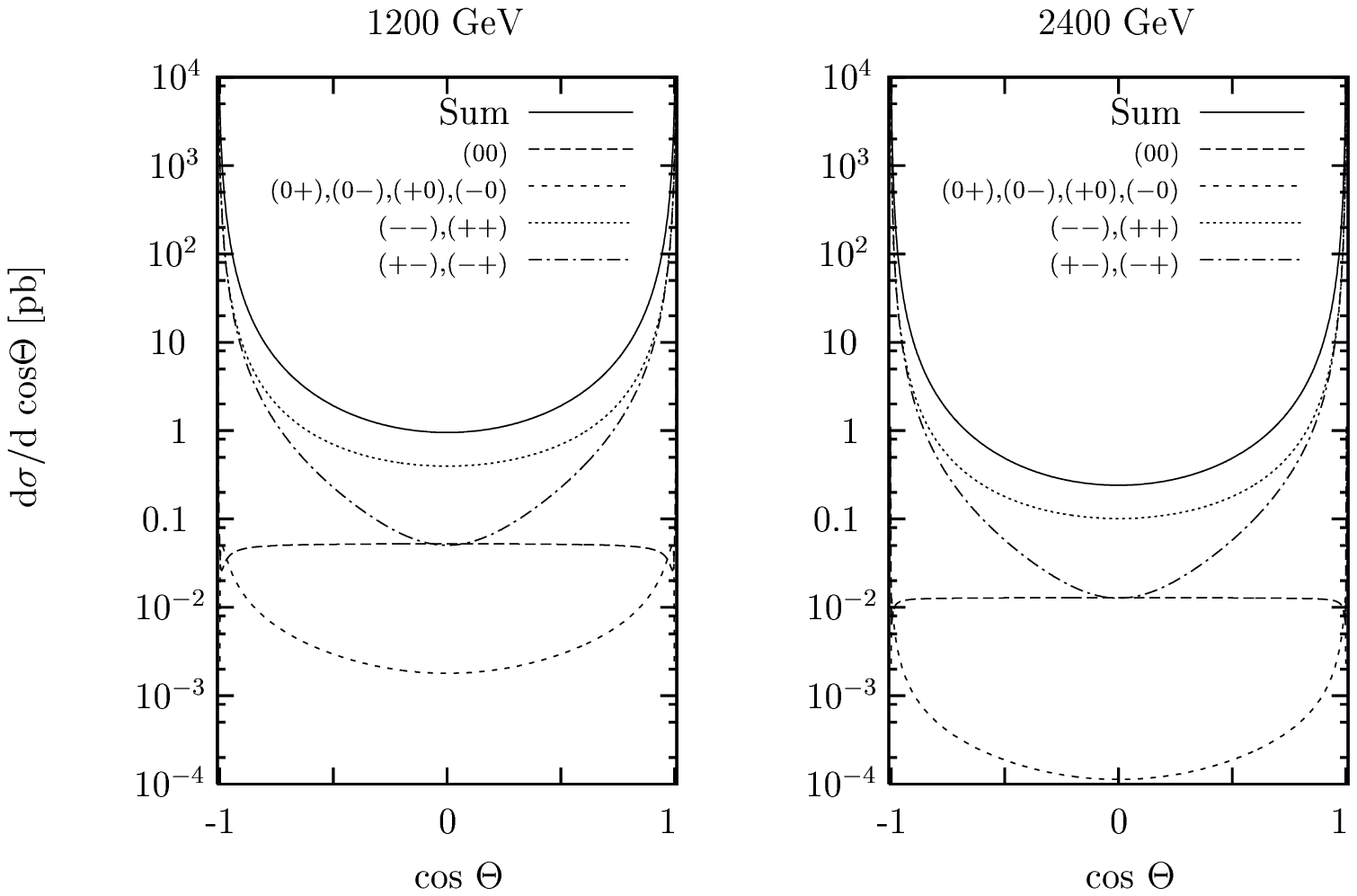}
\caption{\label{fig:undiff}Differential cross section for the process
\mbox{$\gamma \gamma \rightarrow WW$} in the SM with unpolarised photons at
\mbox{$\gamma \gamma$}~c.m.~energies 400~GeV, 640~GeV, 1.2~TeV and 2.4~TeV for
different helicities \mbox{$(\lambda_3, \lambda_4)$} of the $W$~bosons.  For
those curves where more than one helicity combination is indicated the curve
corresponds to a single helicity combination, not to the sum.}
\end{figure}

\begin{figure}
\centering
\includegraphics[totalheight=9cm]{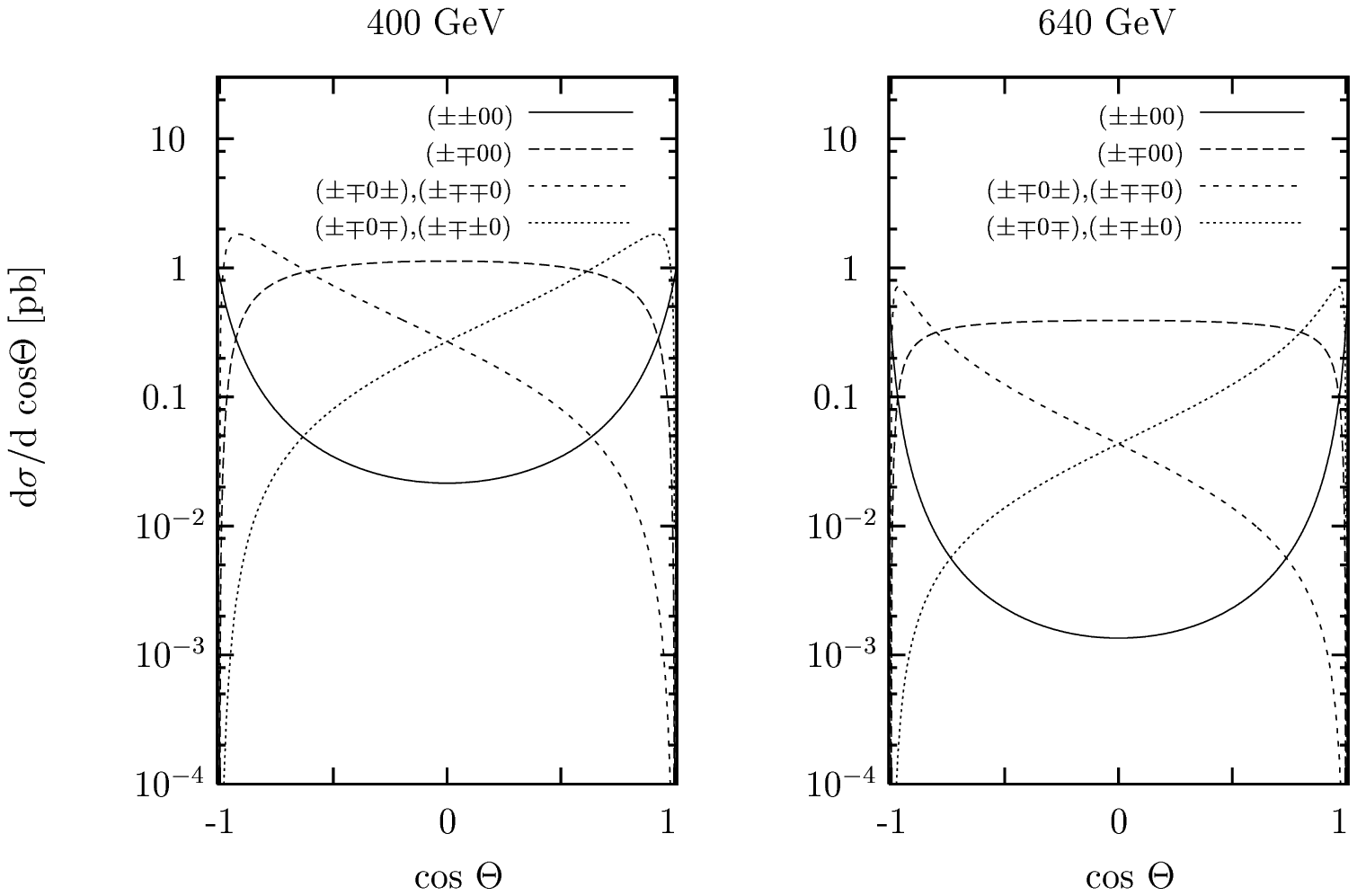}
\includegraphics[totalheight=9cm]{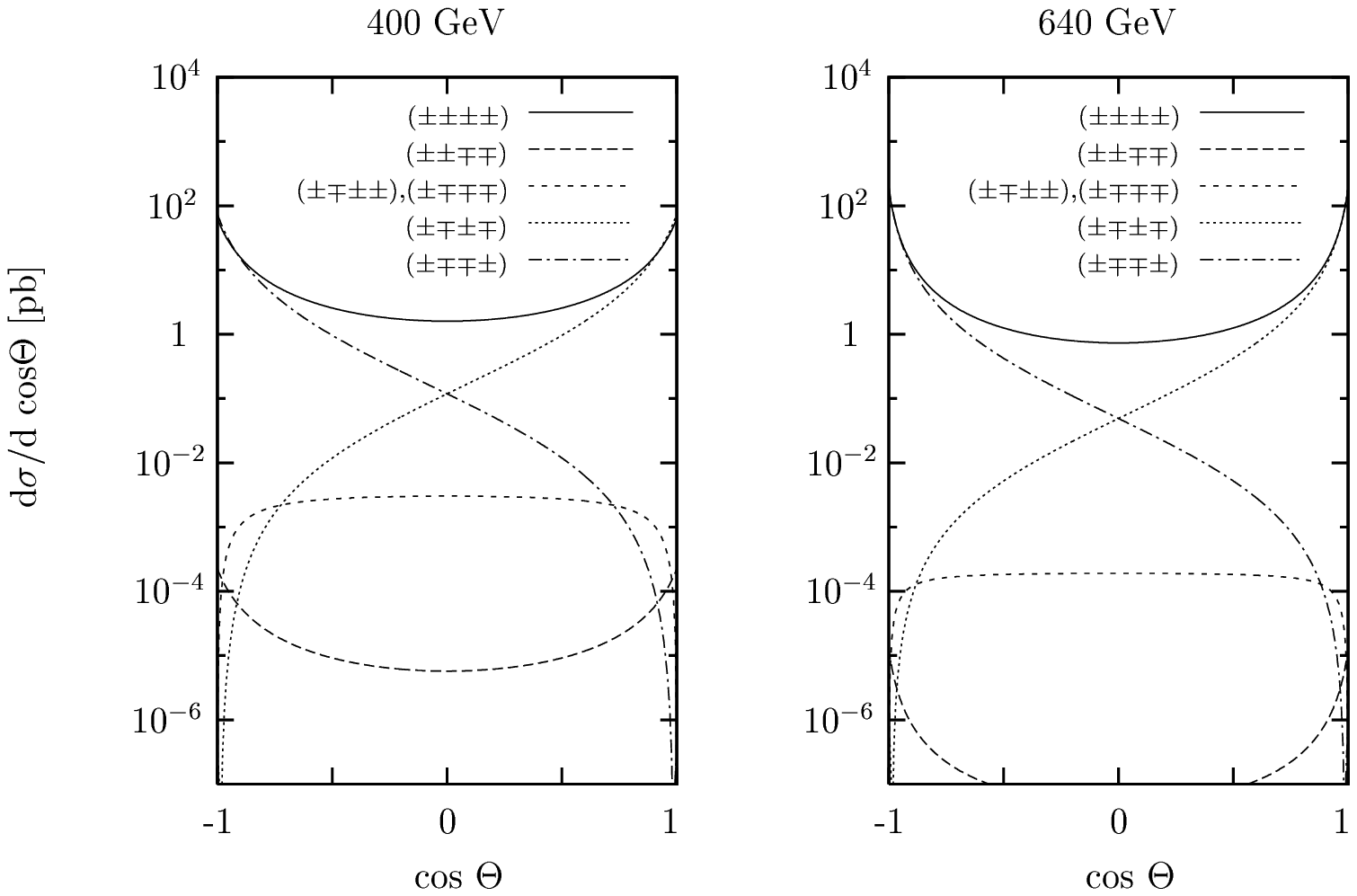}
\caption{\label{fig:diffsig}Differential cross section for the process
\mbox{$\gamma\gamma \rightarrow WW$} in the~SM for different helicities
\mbox{$(\lambda_1, \lambda_2, \lambda_3, \lambda_4)$} of the photons and
$W$~bosons at \mbox{$\gamma \gamma$}~c.m.~energies of 400~GeV and 640~GeV.
The lower part shows the cross section for purely transverse final states,
whereas in the upper part one or both $W$s are longitudinal.  For those curves
where more than one helicity combination is indicated the curve corresponds to
a single helicity combination, not to the sum.}
\end{figure}

We would like to mention some features of the differential
production cross section in the~SM, see Figs.~\ref{fig:undiff}
and~\ref{fig:diffsig}. Its \mbox{$\cos\Theta$}-dependence can be
understood from the conservation of angular momentum in the
\mbox{$\gamma \gamma$}~c.m.~system. Photons with opposite helicities
lead to an initial state with $z$-component of angular momentum $\pm
2$. From this one cannot produce $W^-$ and $W^+$ emitted along the $z$
axis ($\cos \Theta = \pm 1$) with identical helicities, since this
would be a state with $z$ component of angular momentum zero, see
Fig.~\ref{fig:diffsig}.  However, emission at an angle \mbox{$0 <
\Theta <
\pi$} is possible.  In the~SM, two photons with identical helicities can only
produce $W$ bosons with identical helicities.  Furthermore, if the photons
have identical helicities and the $W$~bosons as well, the production of
$W$~bosons with helicity different to that of the photons is suppressed with
rising energy.  Moreover, it is apparent from Figs.~\ref{fig:undiff} and
\ref{fig:diffsig} that in the~SM the bulk of~$W$s is transverse and is emitted
at a small angle to the beam axis, i.e.\ \mbox{$\cos \Theta \approx \pm 1$}.


\begin{figure}
\centering	
 \psfrag{xlab}{\hspace*{-0.15cm}$\cos\Theta$}
 \psfrag{ylab}{\hspace*{-1.2cm}\raisebox{-0.1cm}{$\text{d}\sigma_i/\text{d}\cos\Theta[{\rm pb}]$}}
 \psfrag{h1}[bl][br][0.9]{\hspace*{-0.74cm}$h_W$}
 \psfrag{h3}[bl][br][0.9]{\hspace*{-0.8cm}$h_{W\! B}$}
 \psfrag{h57}[bl][br][0.9]{\hspace*{-0.8cm}$h_{\varphi W\! B}$}
 \psfrag{l3, l4 = 0, 0}{\hspace*{-0.1cm}$(\lambda_3,\lambda_4)=(0,0)$} 
 \psfrag{l3, l4 = 0, -1}{\hspace*{-0.7cm}$(\lambda_3,\lambda_4)=(0,\pm), (\pm,0)$}
 \psfrag{l3, l4 = -1, 1}{\hspace*{-0.1cm}$(\lambda_3,\lambda_4)=(\pm,\mp)$}
 \psfrag{l3, l4 = 1, 1}{\hspace*{-0.1cm}$(\lambda_3,\lambda_4)=(\pm,\pm)$}
\includegraphics[height=18.5cm]{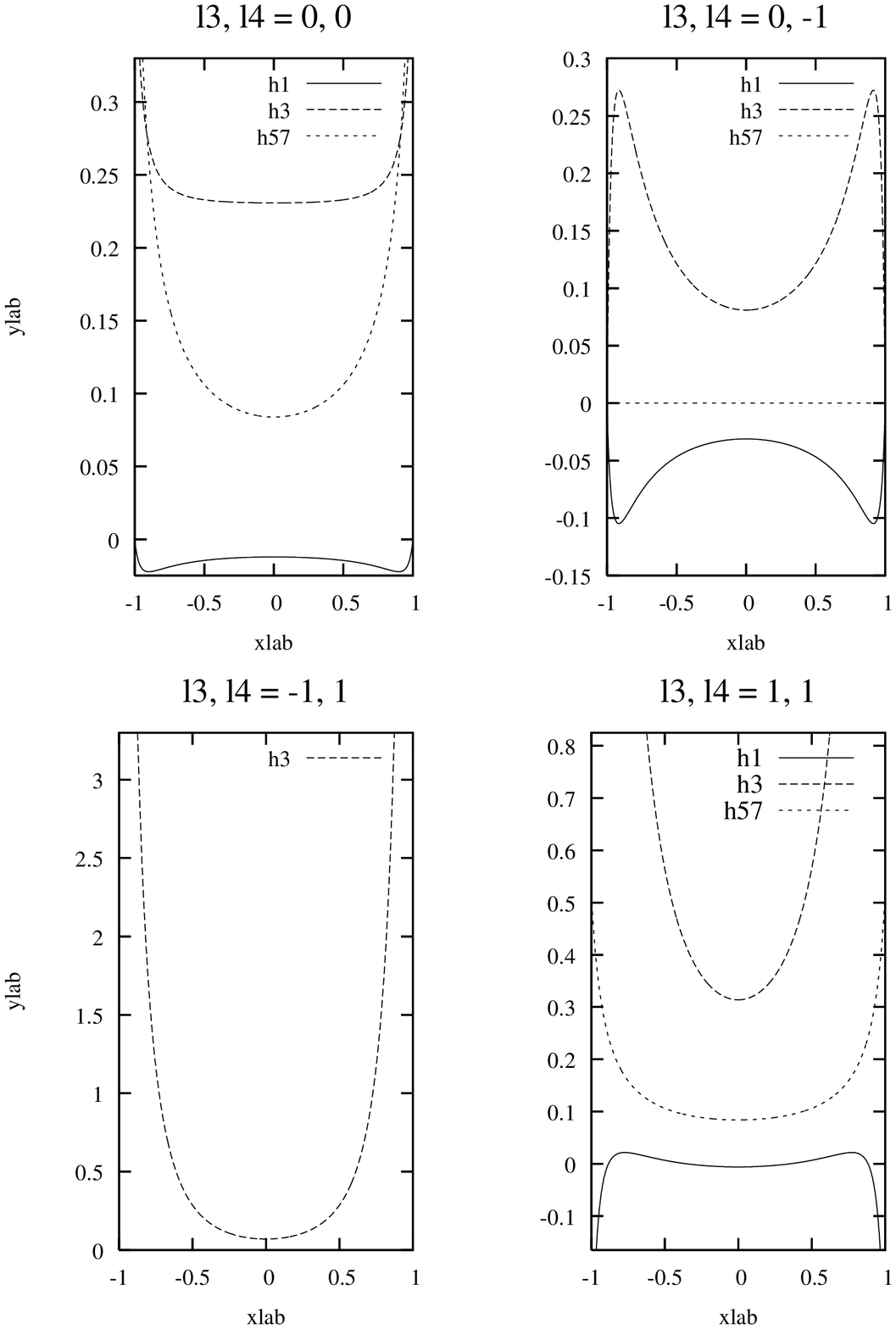}
\caption{\label{fig:undiffh}Anomalous contributions
$\text{d}\sigma_i/\text{d}\cos\Theta$ (\ref{eq-dsigmadecomp}) to the differential
cross section for the process \mbox{$\gamma \gamma \rightarrow WW$}
with unpolarised photons at a
\mbox{$\gamma \gamma$}~c.m.~energy of 400~GeV for
different helicities \mbox{$(\lambda_3, \lambda_4)$} of the
$W$~bosons.}
\end{figure}

We also want to discuss the anomalous contributions to the
differential cross sections (\ref{eq-diffs}) after integration over
$\varphi,\vartheta,\bar{\varphi},\bar{\vartheta}$. Expanding this
cross section in terms of $h_i$,
\begin{equation}
 \frac{\text{d}\sigma}{\text{d}\cos\Theta}=
  \frac{\text{d}\sigma_\text{SM}}{\text{d}\cos\Theta}
+ \sum_i h_i \frac{\text{d}\sigma_i}{\text{d}\cos\Theta} \; + \; O(h^2)\,,
\label{eq-dsigmadecomp}
\end{equation}
one can define the cross section contributions
$\text{d}\sigma_i/\text{d}\cos\Theta$ shown in Fig.~\ref{fig:undiffh}
for the $CP$-conserving couplings. To leading order in the $h_i$ the
$CP$-violating contributions $\text{d}\sigma_i/\text{d}\cos\Theta$
vanish, due to symmetry arguments, see Sect.~\ref{sec-dissymm}. The
shape of the SM and the anomalous cross section contributions varies
significantly, compare Fig.~\ref{fig:undiff} and
Fig.~\ref{fig:undiffh}. Thus it is possible to get some information on
the $CP$-conserving anomalous couplings just from the differential
cross section (\ref{eq-dsigmadecomp}). We get information both on
$CP$-conserving and $CP$-violating couplings if we take the angular
distributions of the final state fermions in (\ref{eq-process}) into
account. In the companion paper \cite{Nachtmann:opti} this will be
done using optimal observables which guarantee the best possible
accuracy in the measurement of the anomalous couplings from a purely
statistical point of view.

\begin{figure}
\centering
 \psfrag{xlab}{\hspace*{.9cm}\raisebox{1ex}{\huge$\sqrt{s}\,[\text{GeV}]$}}
 \psfrag{TITEL}[bc]{}
 \psfrag{ylab}{{\huge$\sigma[{\rm pb}]$}} 
 \psfrag{sm}[bl][br][1.8]{\hspace*{1.3cm}\raisebox{0.4ex}{SM}}
 \psfrag{h1}[bl][br][1.8]{\hspace*{1.3cm}$h_W$}
 \psfrag{h2}[bl][br][1.8]{\hspace*{1.3cm}$h_{\tilde{W}}$}
 \psfrag{h3}[bl][br][1.8]{\hspace*{1.3cm}$h_{W\! B}$}
 \psfrag{h4}[bl][br][1.8]{\hspace*{1.3cm}$h_{\tilde{W}\! B}$}
 \psfrag{h57}[bl][br][1.8]{\hspace*{1.3cm}$h_{\varphi W\! B}$}
 \psfrag{h68}[bl][br][1.8]{\hspace*{1.3cm}$h_{\varphi
 \tilde{W}\!\tilde{B}}$}
 \psfrag{ylab}{\text{\hspace*{-2.7cm}\huge$(\sigma-\sigma_{\text{SM}})/\sigma_{\text{SM}}[\%]$}} 
\includegraphics[angle=-90,width=15.4cm]{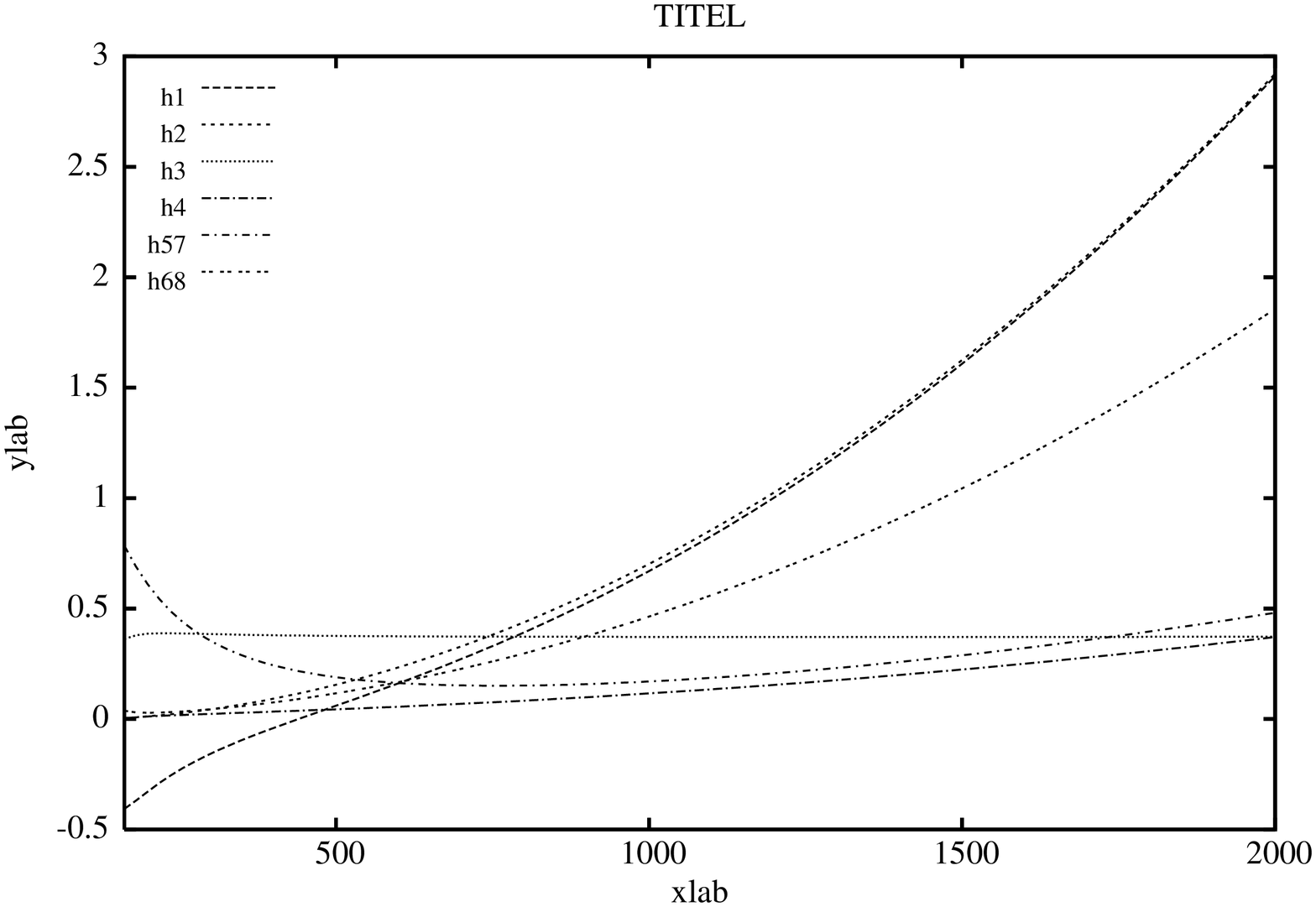}
\caption{\label{fig:totsig}Total cross section for the process
\mbox{$\gamma\gamma \rightarrow WW$}. We show here the unpolarised case,
that is, the average over initial and sum over final-state
helicities. The upper plot shows the total cross section in the SM. In
the lower plot the relative corrections to the SM cross section
including contributions up to quadratic order in the $h_i$ are
shown. For each of these corrections one nonvanishing coupling
contribute whereas all other anomalous couplings are set to zero.  The
values chosen for the couplings are $h_i= 10^{-2}$ for
$i=W,\tilde{W},\tilde{W}\!B,\varphi W\!B,\varphi \tilde{W}\!\tilde{B}$
and $h_i=10^{-3}$ for $i=W\!B$.}
\end{figure}

Finally, we discuss the total cross-section for the process
$\gamma\gamma\to W W$ for unpolarised photons including effects
from anomalous couplings. In Fig.~\ref{fig:totsig} we show the total
SM cross section and the relative corrections
$(\sigma-\sigma_{\text{SM}})/\sigma_{\text{SM}}$ to the total cross
section coming from anomalous contributions up to quadratic order in
the $h_i$. For each of these relative corrections only one coupling
$h_i$ is different from zero. As expected, the anomalous cross
sections rise with the energy if anomalous couplings are taken into
account. For the three $CP$ conserving couplings contributing to the
cross section in linear order we get different contributions for
positive and negative values of the couplings $h_i$. In the limit
$s\gg m_W^2$ only the coupling $h_{W\!B}$ contributes with a linear
term in $\sigma$, for all other couplings $h_i$ the anomalous
contributions to the total cross section are in this limit of the
order $h_i^2$, due to the fact that the linear terms vanish after the
integration over the angles related to the final fermion states. We
see from Fig.~\ref{fig:totsig} that for $h_{W\!B}=10^{-3}$, the
maximal value allowed from the analysis of precision observables (see
table 8 of
\cite{Nachtmann:2004ug}), the total cross section at
$\sqrt{s}=500\;\text{GeV}$ is modified only by $\approx 0.4\%$. The
other couplings have similarly small effects already for
$h_i=10^{-2}$. In addition, effects in the cross section are, of
course, not specific to any of the couplings. On the other hand we
will show below that in the complete differential cross section
(\ref{eq-diffs}) the anomalous contributions are dominated by the
terms of the order $h_i$ for reasonable energies. Typical bounds on
anomalous couplings obtainable at an $e^+e^-$ collider at
$\sqrt{s}=500\;\text{GeV}$ are less than $10^{-3}$ with optimal
observables. Quadratic terms of anomalous couplings can then safely be
neglected. For the total cross sections for example they would then be
less then order $10^{-4}$.

\section{\boldm{CP} symmetry}
\label{sec-dissymm}
\setcounter{equation}{0}

To determine the sensitivity to the anomalous couplings---in
particular within the framework of optimal observables considered
in~\cite{Nachtmann:opti}---it is convenient to make use of the
transformation properties of the amplitudes under the combined
discrete symmetry~$CP$ of charge conjugation and parity reversal.
Since our effective Lagrangian does not modify the SM~couplings of
$W$~bosons to fermions these interactions are $CP$~invariant as in
the~SM neglecting the phase in the CKM matrix.  This phase plays no
role in our considerations below. However the effective
Lagrangian~(\ref{eq-Leff}) as a whole is not $CP$~conserving since it
contains CP violating anomalous gauge-boson couplings. The primed
fields of the effective Lagrangian, see App.~\ref{sec-conv}, transform
under~$CP$ as follows:
\begin{align}
\label{eq-cp1}
A^{\prime\,\mu} (x)\; & \rightarrow\;  - A^\prime_{\mu} (x^\prime)\,,\\
\label{eq-cp2}
Z^{\prime\,\mu} (x)\; & \rightarrow \; - Z^\prime_{\mu} (x^\prime)\,,\\
\label{eq-cp3}
W^{\prime \pm \, \mu} (x)\; & \rightarrow \; - W^{\prime \, \mp}_{\mu} (x^\prime)\,,
\end{align}
with \mbox{$x = (x^0, {\bf x})$} and \mbox{$x^\prime = (x^0, -{\bf x})$}.
Without the dimension-six operators in~(\ref{eq-Leff}) the primed
gauge fields are the physical fields.  The Lagrangian is then the SM
Lagrangian, which is invariant under the
transformations~(\ref{eq-cp1}) to~(\ref{eq-cp3}) and an appropriate
choice for the $CP$ transformation of the other SM~fields if the phase
of the CKM~matrix is set to zero. Considering the
full Lagrangian~(\ref{eq-Leff}), the physical fields, which are linear
combinations of the primed fields as given in~\cite{Nachtmann:2004ug},
transform as
\begin{align}
A^{\mu} (x)\; & \rightarrow \; - A_{\mu} (x^\prime)\,,\\
Z^{\mu} (x)\; & \rightarrow \; - Z_{\mu} (x^\prime)\,,\\
W^{\pm \, \mu} (x)\; & \rightarrow \; - W^{\mp}_{\mu} (x^\prime)\,.
\end{align}
This implies the following relations for the different parts of the
amplitudes in~(\ref{eq-Mdecomp}) 
\begin{align}
\label{eq-amplcp1}
\mathcal{M}_{\rm SM} (\lambda_1,\lambda_2;\lambda_3,\lambda_4) & = 
(-1)^{(\lambda_3 + \lambda_4)} 
\mathcal{M}_{\rm SM} (-\lambda_2,-\lambda_1;-\lambda_4,-\lambda_3)\,,
\\
\label{eq-amplcp2}
\mathcal{M}_i (\lambda_1,\lambda_2;\lambda_3,\lambda_4) & = 
\pi_i\;(-1)^{(\lambda_3 + \lambda_4)}
\mathcal{M}_i (-\lambda_2,-\lambda_1;-\lambda_4,-\lambda_3)
\end{align}
with \mbox{$\pi_i = +1$} for the couplings $h_W$, $h_{\varphi W}$, $h_{\varphi
B}$, $h_{W\! B}$, and \mbox{$\pi_i = -1$} for $h_{\tilde{W}}$,
$h_{\varphi \tilde{W}}$, $h_{\varphi \tilde{B}}$ and $h_{\tilde{W}\!
B}$.  Particle momenta are understood to be equal on both sides
of~(\ref{eq-amplcp1}) and~(\ref{eq-amplcp2}).  From these two equations we can
see that the couplings $h_W$, $h_{\varphi W}$, $h_{\varphi B}$,
$h_{W\! B}$ are $CP$~conserving---as are $h_{\varphi}^{(1)}$ and
$h_{\varphi}^{(3)}$---whereas the couplings $h_{\tilde{W}}$, $h_{\varphi
\tilde{W}}$, $h_{\varphi \tilde{B}}$ and $h_{\tilde{W}\! B}$ are
$CP$~violating.

\begin{figure}
\centering
\includegraphics[totalheight=2.5cm]{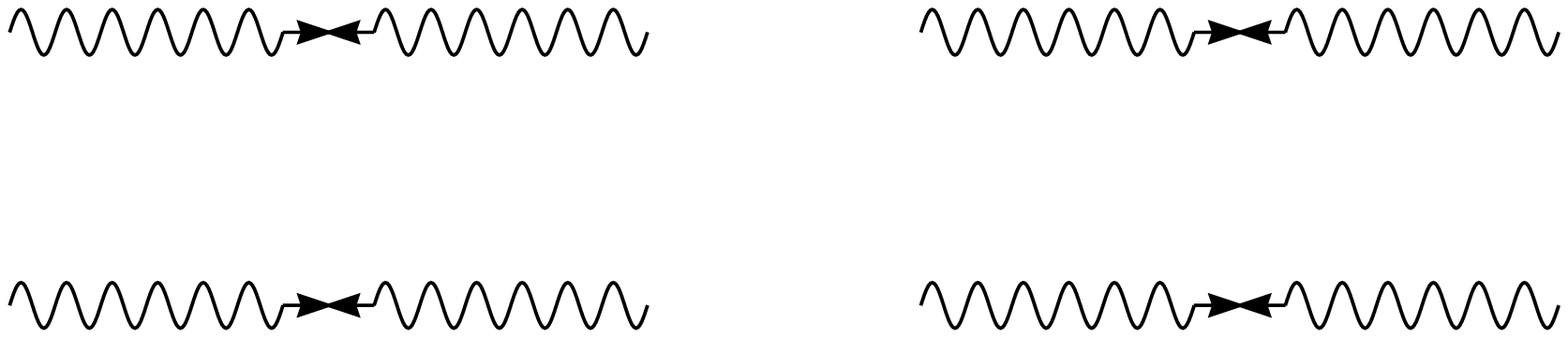}
\setlength{\unitlength}{1cm}
\begin{picture}(5,1)
\put(2.15,3.1){$\longleftrightarrow$}
\put(2.15,1.2){$\longleftrightarrow$}
\put(2.2,3.6){$CP$}
\put(2.2,1.7){$CP$}
\put(-1.9,3.6){$\gamma$}
\put(0.5,3.6){$\gamma$}
\put(4.3,3.6){$\gamma$}
\put(6.6,3.6){$\gamma$}
\put(-1.9,1.7){$\gamma$}
\put(0.5,1.7){$\gamma$}
\put(4.3,1.7){$\gamma$}
\put(6.6,1.7){$\gamma$}
\put(-2.1,2.6){$\Longrightarrow$}
\put(0.3,2.6){$\Longrightarrow$}
\put(4,2.6){$\Longrightarrow$}
\put(6.5,2.6){$\Longrightarrow$}
\put(-2.1,0.7){$\Longrightarrow$}
\put(0.3,0.7){$\Longleftarrow$}
\put(4,0.7){$\Longleftarrow$}
\put(6.5,0.7){$\Longrightarrow$}
\end{picture}
\caption{\label{fig:cpdiag}Behaviour of a \mbox{$\gamma \gamma$}~state with
opposite (top) and same (bottom) helicities under a $CP$~transformation.}
\end{figure}

Under the condition that the initial state, phase-space cuts and detector
acceptance are invariant under a $CP$~transformation, a $CP$~odd observable
gets a nonzero expectation value only if $CP$ is violated in the interaction.
Consider now the reaction \mbox{$\gamma \gamma \rightarrow WW$} for fixed
photon~energies in the c.m.~system.  The initial state is invariant under $CP$
for unpolarised photon beams as well as for the states with \mbox{$|J_z| =
2$}, where $J_z$ is the total angular momentum along the $z$-axis, see
Fig.~\ref{fig:cpdiag}.  For a photon collider with identical energy spectra
for the two photons the same $CP$~invariance properties hold.
This will be exploited for our numerical calculations in~\cite{Nachtmann:opti}
assuming unpolarised photons, so that we indeed have a $CP$~invariant initial
state.  There to linear order in the $h_i$ the expectation values of $CP$~even (odd)
optimal observables only involve the $CP$~conserving (violating) couplings.
Then---provided couplings with the same discrete symmetry properties are grouped
together---the covariance matrix of the optimal observables is block diagonal
with two blocks corresponding to the $CP$~conserving and $CP$~violating
couplings, respectively.

\section{Conclusions}
\label{sec-conc}
\setcounter{equation}{0}

We have analysed the phenomenology of the gauge-boson sector of an
electroweak effective Lagrangian that is locally $SU(2)\times U(1)$
invariant.  In addition to the SM~Lagrangian we include all ten
dimension-six operators that are built either only from the
gauge-boson fields of the~SM or from the gauge-boson fields combined
with the SM-Higgs field. 

We have investigated in detail the effects of the effective Lagrangian
on the reaction \mbox{$\gamma \gamma \rightarrow WW$} in the
photon-collider mode at an ILC.  In this process the three
$CP$~conserving couplings $h_W, h_{W\! B}, h_{\varphi W\! B}$ and the three
$CP$~violating couplings $h_{\tilde{W}}, h_{\tilde{W}\! B}, h_{\varphi
\tilde{W}\! \tilde{B}}$ are measurable, whereas the remaining three $CP$~conserving
ones $h_\varphi^{(1)}, h_\varphi^{(3)}, h^\prime_{\varphi W\! B}$ and
a $CP$~violating one $h^\prime_{\varphi
\tilde{W}\!\tilde{B}}$ are not measurable, see Sect.~\ref{sec-cross}.  The
strongest possible bounds on the anomalous couplings that can be
obtained in our reaction will be computed in the companion
paper~\cite{Nachtmann:opti} by means of optimal observables.  There
are two $CP$~conserving and two $CP$~violating couplings occurring
both in \mbox{$\gamma \gamma \rightarrow WW$} and in \mbox{$e^+e^-
\rightarrow WW$}, whereas the couplings $h_{\varphi W\! B}$ and $h_{\varphi
\tilde{W}\! \tilde{B}}$ are only measurable in $\gamma\gamma\to WW$ .
A comparison of the sensitivity to the anomalous couplings achievable
in the two reactions will be performed in~\cite{Nachtmann:opti}. Such
a quantitative analysis is important to decide how much total
luminosity is required in each mode of an ILC.  As already
explained in Sect.~\ref{sec-intro}, our approach, using the effective
Lagrangian (\ref{eq-Leff}) instead of form factors, is perfectly
suited for a comprehensive study of all constraints on the $h_i$
coming from different modes at an ILC and from high precision
observables. We have seen that in any case the
\mbox{$e^+ e^-$} and the \mbox{$\gamma \gamma$} modes deliver complementary constraints on the
anomalous couplings of the effective Lagrangian considered.  Both
modes are indispensable for a comprehensive study of the gauge-boson
sector at an ILC.


\section*{Acknowledgements}

The authors are grateful to M.~Diehl for reading a draft of this
manuscript and to A.~Denner and A.~de~Roeck for useful discussions.
This work was supported by the German Bundesministerium f\"ur Bildung
und Forschung, BMBF project no. 05HT4VHA/0, and the Deutsche
Forschungsgemeinschaft through the Graduierten\-kolleg ``Physikalische
Systeme mit vielen Freiheitsgraden''.


\begin{appendix}

\section{Effective Lagrangian}
\label{sec-conv}
\setcounter{equation}{0}

In this appendix we give the definition of the effective
Lagrangian~(\ref{eq-Leff}).  For the SM Lagrangian~$\mathscr{L}_{0}$
we use the conventions of~\cite{Nachtmann:ta}.  Restricting ourselves
to the electroweak interactions and neglecting neutrino masses
$\mathscr{L}_0$ is given by
\begin{equation}
\begin{split}
\label{eq-smlagr}
\mathscr{L}_0  = &
- \frac{1}{4} W^i_{\mu \nu} W^{i \, \mu \nu} 
- \frac{1}{4} B_{\mu \nu} B^{\mu \nu}
+ \left(\mathcal{D}_{\mu} \varphi \right)^{\dag} \left(\mathcal{D}^{\mu}
  \varphi\right)
+ \mu^2 \varphi^{\dag} \varphi - \lambda \left(
\varphi^{\dag} \varphi \right)^2 	
\\[1ex]
&  + i \overline{L}  \capslash{\mathcal{D}} L 
+ i \overline{E} \capslash{\mathcal{D}} E
+ i \overline{Q} \capslash{\mathcal{D}} Q 
+ i \overline{U} \capslash{\mathcal{D}} U 
+ i \overline{D} \capslash{\mathcal{D}} D 
\\[1ex]
&  - \left(\overline{E}\, \Gamma_E \, \varphi^{\dag} L 
+ \overline{U}\, \Gamma_U \, \tilde{\varphi}^{\dag} Q
+ \overline{D}\, \Gamma_D \, \varphi^{\dag} Q + {\rm H.c.}
\right)\,.
\end{split}
\end{equation}
The \mbox{$3\times 3$}-Yukawa matrices have the form
\begin{align}
\Gamma_{E} & =  {\rm diag} (c_{\rm e}, c_{\mu}, c_{\tau}),\\
\Gamma_{U} & =  {\rm diag} (c_{\rm u}, c_{\rm c}, c_{\rm t}),\\
\Gamma_{D} & =  V {\rm diag} (c_{\rm d}, c_{\rm s}, c_{\rm b}) V^{\dag},
\end{align}
where the diagonal elements all obey \mbox{$c_i \geq 0$} and $V$ is the
CKM~matrix.  With these conventions the matrices $\Gamma_E$, $\Gamma_U$,
$\Gamma_D$ correspond to the matrices $C_{\ell}$, $C'_q$, $C_q$
in~\cite{Nachtmann:ta}, respectively.  The vector of the three left-handed
lepton doublets is denoted by~$L$, of the right-handed charged leptons by~$E$,
of the left-handed quark doublets by~$Q$, and of the right-handed up- and
down-type quarks by~$U$ and~$D$.  The Higgs field is denoted by~$\varphi$.
After electroweak symmetry breaking we can choose the form
\begin{equation}
\label{eq-Higgs}
\varphi(x) = \frac{1}{\sqrt{2}} \left(\begin{matrix} 0 \\ 
v + H^\prime(x)\end{matrix}\right)\,,
\end{equation}
where $v$ is the vacuum expectation value and \mbox{$H'(x)$} is a real scalar
boson field.  With the full Lagrangian~$\mathscr{L}_{\rm eff}$
(\ref{eq-Leff}) it is related
to the physical Higgs-boson field \mbox{$H(x)$} through
\be
\label{eq-Higgsren}
H = \sqrt{1 + \left(h_{\varphi}^{(1)} + h_{\varphi}^{(3)}\right)\! /2} \;
H^\prime\,.
\ee
The factor depending on the anomalous couplings stems from
the renormalisation of the Higgs field, see~\cite{Nachtmann:2004ug}.
In~(\ref{eq-smlagr}) we have used the definitions
\be
\tilde{\varphi} = \varepsilon \varphi^*\,,\msp \msp
\varepsilon = \left(\begin{matrix} 0 & 1 \\ -1 & 0 \end{matrix}\right)\,,
\ee
and for the covariant derivative 
\be
\label{eq-covderiva}
\mathcal{D}_{\mu} = \partial_{\mu} + i g  W^i_{\mu} \mathbf{T}_i + i g' B_{\mu}
\mathbf{Y}\,,
\ee
where $\mathbf{T}_i$ and $\mathbf{Y}$ are the generating operators of
weak-isospin and weak-hypercharge transformations.  For the left-handed
fermion fields and the Higgs doublet we have \mbox{$\mathbf{T}_i = \tau_i /
2$}, where $\tau_i$ are the Pauli matrices.  For the right-handed fermion
fields we have \mbox{$\mathbf{T}_i = 0$}.  The hypercharges of the fermions
and of the Higgs doublet are listed in Tab.~\ref{tab:hyper}.
\begin{table}
\centering
\begin{tabular}{cccccccc}
\hline
 &&&&&&&\\[-.45cm]
 & $L$ & $E$ & $Q$ & $U$ & $D$ & $\varphi$\\
\hline
 &&&&&&&\\[-.4cm]
$y$ & $-\frac{1}{2}$ & $-1$ & $\frac{1}{6}$ & $\frac{2}{3}$ & $-\frac{1}{3}$ &
$\frac{1}{2}$\\[.1cm]
\hline
\end{tabular}
\caption{\label{tab:hyper}Weak hypercharges of the fermions and of the Higgs
 doublet.}
\end{table}
The field strengths are
\be
\label{eq-gifs}
W^i_{\mu \nu} = \partial_{\mu}W^i_{\nu} - \partial_{\nu}W^i_{\mu}
- g \, \epsilon_{ijk} \, W^j_{\mu} W^k_{\nu}\,,\msp \msp
B_{\mu \nu} = \partial_{\mu}B_{\nu} -
\partial_{\nu}B_{\mu}\,.
\ee
Note that the signs in front of the gauge couplings in~(\ref{eq-covderiva})
and~(\ref{eq-gifs}) differ from the conventions of~\cite{Buchmuller:1985jz},
which leads to sign changes in some of the dimension-six operators listed
below.  The dimension-six operators of~$\mathscr{L}_{2}$ in~(\ref{eq-dim6op})
are defined as follows, see (3.5), (3.6), and (3.41) to~(3.44)
in~\cite{Buchmuller:1985jz},
\begin{alignat}{4}
\label{eq-Ow}
&O_W && = \epsilon_{ijk} \, W^{i \, \nu}_{\mu} W^{j \, \lambda}_{\nu}
W^{k \, \mu}_{\lambda}\,,\phantom{\frac{1}{2}}&\;\;\;\;\;\;\;\;\;\;
&O_{\tilde{W}} && = \epsilon_{ijk} \, \tilde{W}^{i \, \nu} _{\mu} W^{j
\, \lambda} _{\nu} W^{k \, \mu}_{\lambda}\,,\\[.1cm]
\label{eq-Ophiw}
&O_{\varphi W} && = \frac{1}{2} \left( \varphi^{\dag}\varphi
\right)\,W^i_{\mu \nu} W^{i\, \mu \nu}\,,&\;\;\;\;\;\;\;\;\;\;
&O_{\varphi \tilde{W}} && = \left( \varphi^{\dag} \varphi
\right)\,\tilde{W}^i_{\mu \nu} W^{i \, \mu \nu}\,,\\[.1cm]
\label{eq-Ophib}
&O_{\varphi B} && = \frac{1}{2} \left( \varphi^{\dag}\varphi
\right)\,B_{\mu \nu} B^{\mu \nu}\,,&\;\;\;\;\;\;\;\;\;\;
&O_{\varphi \tilde{B}} && = \left( \varphi^{\dag}\varphi
\right)\,\tilde{B}_{\mu \nu} B^{\mu \nu}\,,\\[.1cm]
\label{eq-Owb}
&O_{W\! B} && = \left( \varphi^{\dag} \tau^i \varphi
\right)\,W^i_{\mu \nu}B^{\mu \nu}\,,\phantom{\frac{1}{2}}&\;\;\;\;\;\;\;\;\;\;
&O_{\tilde{W}\! B} && = \left( \varphi^{\dag} \tau^i \varphi
\right)\,\tilde{W}^i_{\mu \nu} B^{\mu \nu}\,,\\[.1cm]
\label{eq-Ophi1}
&O_{\varphi}^{(1)} && = \left( \varphi^{\dag}\varphi \right) \left(
\mathcal{D}_{\mu} \varphi \right)^{\dag} \left(\mathcal{D}^{\mu} \varphi
\right)\,,\phantom{\frac{1}{2}}&\;\;\;\;\;\;\;\;\;\;
&O_{\varphi}^{(3)} && = \left(\varphi^{\dag} \mathcal{D}_{\mu} \varphi
\right)^{\dagger}\left( \varphi^{\dag} \mathcal{D}^{\mu} \varphi \right)\,.
\end{alignat}
The dual field strengths are
\be
\tilde{W}^i_{\mu \nu} = \frac{1}{2} \epsilon_{\mu\nu\rho\sigma}
W^{i\,\rho\sigma}\,,\msp \msp
\tilde{B}_{\mu \nu} = \frac{1}{2} \epsilon_{\mu\nu\rho\sigma}
B^{\rho\sigma}\,.
\ee
In \cite{Nachtmann:2004ug} the original $W_{\mu}^i$ and
$B_{\mu}$ fields in~$\mathscr{L}_{\rm eff}$ are expressed in terms of the
primed fields $Z'$, $A'$ and~$W^{\prime \pm}$ as follows:
\begin{align}
\label{eq-prfielddef1}
W_{\mu}^1 &=\frac{1}{\sqrt{2}} 
\left(W_{\mu}^{\prime +} + W_{\mu}^{\prime -}\right)\,,
\qquad 
& W_{\mu}^2 & =\frac{i}{\sqrt{2}}
\left(W_{\mu}^{\prime +}-W_{\mu}^{\prime -}\right)\,,\\
\label{eq-prfielddef2}
W^3_{\mu}& = c_{\rm w}' \, Z'_{\mu} + s_{\rm w}' \,  A^\prime_{\mu}\,,
\qquad 
& B_{\mu} &= - s_{\rm w}' \, Z'_{\mu} + c_{\rm w}' \, A^\prime_{\mu}\,,
\end{align}
where
\begin{alignat}{3}
\label{eq-sinepr}
&s'_{\rm w} \; & \equiv & \; \sin \theta'_{\rm w} \; & = & \;
      \frac{g'}{\sqrt{g^2 + g^{\prime \, 2}}}\,,\\
\label{eq-cosinepr} 
&c'_{\rm w} \; & \equiv & \; \cos \theta'_{\rm w} \; & = & \;
      \frac{g}{\sqrt{g^2 + g^{\prime \, 2}}}
\end{alignat}
are the sine and cosine of the weak mixing angle in the~SM, determined by the
\mbox{$SU(2)$} and \mbox{$U(1)_Y$} couplings $g$ and $g^\prime$
of~$\mathscr{L}_{0}$. The primed fields are the physical gauge-boson
fields in absence of anomalous couplings.  With anomalous couplings
the $W$-boson field must be renormalised by a factor that depends on
the~$h_i$, similarly to the Higgs-boson field in~(\ref{eq-Higgsren}).
The physical neutral gauge-boson fields $A$ and~$Z$ are linear
combinations of $A'$ and~$Z'$ with coefficients depending on the
anomalous couplings.  For these relations we refer to Sect.~3
of~\cite{Nachtmann:2004ug}.


\section{Feynman rules}
\label{sec-feynm}
\setcounter{equation}{0}

\begin{figure}
\hspace{3cm} 
\parbox{5cm}{\includegraphics[totalheight=3.5cm]{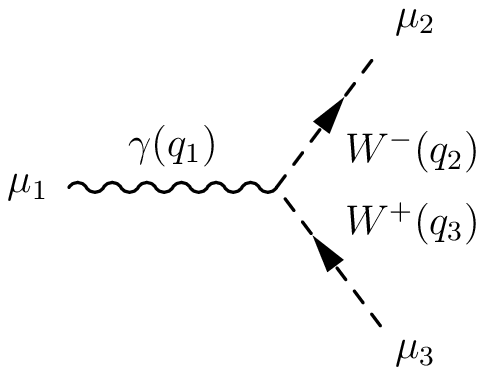}}
$=$
\hspace{0.5cm}
$i e \; \Gamma_{\rm SM}^{(3) \;\mu_1 \mu_2 \mu_3}$ \\
\vspace{1cm}

\hspace{3cm} 
\parbox{5cm}{\includegraphics[totalheight=3.5cm]{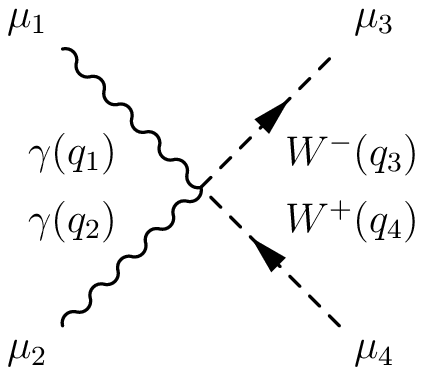}}
$=$
\hspace{0.5cm}
$i e^2 \; \Gamma_{\rm SM}^{(4) \;\mu_1 \mu_2 \mu_3 \mu_4}$\\
\vspace{1cm}

\hspace{3cm} 
\parbox{5cm}{\includegraphics[totalheight=3.5cm]{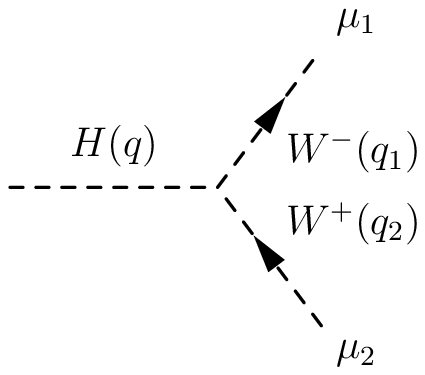}}
$=$
\hspace{0.5cm}
$i \; \Gamma_{\rm SM}^{(WWH) \;\mu_1 \mu_2}$\\
\caption{\label{fig:smfeynrules}Feynman rules from the SM
Lagrangian. The momenta of the particles are indicated in
bracket. The momentum flow is always from left to right.}
\end{figure}

\begin{figure}
\hspace{3cm}
\parbox{5cm}{\includegraphics[totalheight=3.5cm]{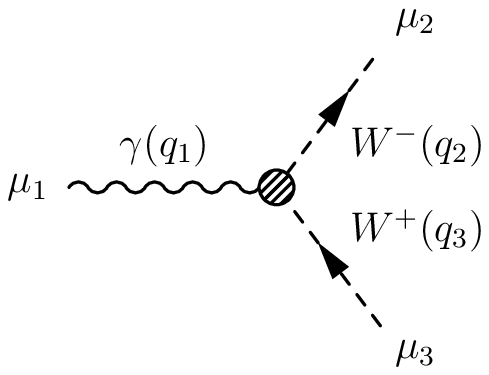}}
$=$
\hspace{0.5cm}
$i e \; \Gamma_{\rm an}^{(3) \;\mu_1 \mu_2 \mu_3}$ 
\vspace{1cm}

\hspace{3cm}
\parbox{5cm}{\includegraphics[totalheight=3.5cm]{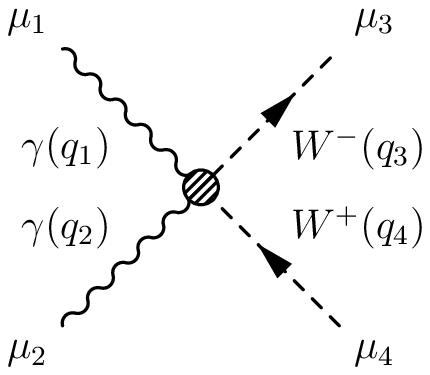}}
$=$
\hspace{0.5cm}
$i e^2 \; \Gamma_{\rm an}^{(4) \;\mu_1 \mu_2 \mu_3 \mu_4}$
\vspace{1cm}

\hspace{3cm}
\parbox{5cm}{\includegraphics[totalheight=3.5cm]{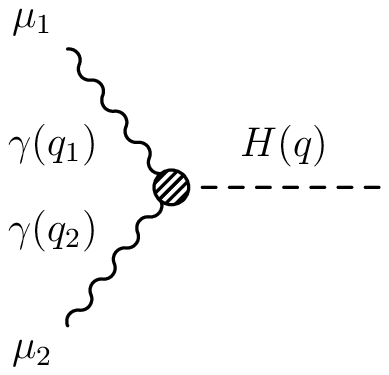}}
$=$
\hspace{0.5cm}
$i \; \Gamma_{\rm an}^{(\gamma \gamma H) \mu_1 \mu_2} $
\caption{\label{fig:anfeynrules}Feynman rules for the anomalous
vertices. The momenta of the particles are indicated in brackets. The
momentum flow is always from left to right.}
\end{figure}

In this section we list the Feynman rules that are necessary for the
evaluation of the diagrams in Figs.~\ref{fig:feyndiag}
to~\ref{fig:anHdiag} in the framework of the effective
Lagrangian~(\ref{eq-Leff}).  These Feynman rules are obtained after
expressing the field operators in~(\ref{eq-Leff}) in the physical
fields $A$, $Z$, $W^{\pm}$ and~$H$, see~\cite{Nachtmann:2004ug}.  All
constants are expressed in terms of the parameters of the
$P_W$~scheme, i.e.\ in terms of \mbox{$\alpha(m_Z)$}, $G_{\rm F}$,
$m_W$, $m_H$ and the ten anomalous couplings~$h_i$, cf.\ Tab.~3
in~\cite{Nachtmann:2004ug}. As in~\cite{Nachtmann:2004ug} we use the
abbreviation \mbox{$e =\sqrt{4 \pi \alpha(m_Z)}$} for the positron
charge at the scale of the $Z$~mass.  After linearisation in the~$h_i$
a vertex with given fields can be written as the sum of the SM~vertex
and the vertices proportional to the~$h_i$.  Here we list the
SM~vertices and the vertices proportional to the~$h_i$ separately.  In
all cases the momenta belonging to lines to the left of a vertex are
incoming whereas those belonging to lines to the right of a vertex are
outgoing. The Feynman rules for the SM~vertices and for the anomalous
\mbox{$\gamma WW$}, \mbox{$\gamma \gamma WW$} and 
\mbox{$\gamma \gamma H$} vertices are shown in
Fig.~\ref{fig:smfeynrules} and Fig.~\ref{fig:anfeynrules},
respectively.

The SM vertex functions are
\begin{align}
\Gamma_{\rm SM}^{(3) \;\mu_1 \mu_2 \mu_3 } & =  g^{\mu_1 \mu_2} 
(q_1+q_2)^{\mu_3}- g^{\mu_1 \mu_3} (q_1+q_3)^{\mu_2} + g^{\mu_2 \mu_3} 
(q_3-q_2)^{\mu_1}\,,
\\
\Gamma_{\rm SM}^{(4) \;\mu_1 \mu_2 \mu_3 \mu_4} & =  g^{\mu_1 \mu_3} 
g^{\mu_2 \mu_4}+ g^{\mu_1 \mu_4} g^{\mu_2 \mu_3}- 2 g^{\mu_1 \mu_2} 
g^{\mu_3 \mu_4}\,,
\\
\Gamma_{\rm SM}^{(WWH) \;\mu_1 \mu_2} & =  2 m_W^2
\left(\sqrt{2} G_{\rm F}\right)^{1/2} g^{\mu_1 \mu_2}\,.
\end{align}
\clearpage

\noindent
The anomalous vertex functions are (for
the definition of $s_1$ and $c_1$ see Eq.~(\ref{s1c1})): 
\begin{align}
\begin{split}
\Gamma_{\rm an}^{(3) \;\mu_1 \mu_2 \mu_3 } & = 6 h_W\,\sqrt{2}G_\text{F}\,
\frac{s_1}{e} \Big(q_2^{\mu_1} q_3^{\mu_2} q_1^{\mu_3} - q_3^{\mu_1}
q_1^{\mu_2} q_2^{\mu_3} 
\\  
& \hphantom{=6 h_W\,\sqrt{2}G_\text{F}\,\frac{s_1}{e} \Big(}	
 + (g^{\mu_2 \mu_3} q_3^{\mu_1} - g^{\mu_1 \mu_3}
q_3^{\mu_2})\: q_1\!\cdot\! q_2  
\\
& \hphantom{=6 h_W\,\sqrt{2}G_\text{F}\,\frac{s_1}{e} \Big(}	
+ (g^{\mu_1 \mu_2} q_2^{\mu_3} - g^{\mu_2 \mu_3}q_2^{\mu_1})\:
q_1\!\cdot\! q_3 
\\
& \hphantom{=6 h_W\,\sqrt{2}G_\text{F}\,\frac{s_1}{e} \Big(}	
+ (g^{\mu_1 \mu_3} q_1^{\mu_2} - g^{\mu_1 \mu_2}
q_1^{\mu_3})\: q_2\!\cdot\! q_3 \Big)  
\\
& \quad
+ 6 h_{\tilde{W}}\,\sqrt{2}G_\text{F}\,\frac{s_1}{e} \Big(
q_3^{\mu_2} \epsilon^{\mu_1 \mu_3 \rho \sigma} q_{1 \, \rho} q_{2 \, \sigma} - 
q_2^{\mu_3} \epsilon^{\mu_1 \mu_2 \rho \sigma} q_{1 \, \rho} q_{3 \, \sigma}  
\\
&
\quad
\hphantom{+ 6 h_{\tilde{W}}\,\sqrt{2}G_\text{F}\,\frac{s_1}{e}\Big(}
  - g^{\mu_2 \mu_3} \epsilon^{\mu_1 \nu \rho \sigma} q_{1
  \,\nu} q_{2\, \rho} q_{3\, \sigma} - q_2\!\cdot\! q_3\:\epsilon^{\mu_1 \mu_2 \mu_3
  \sigma} q_{1 \, \sigma} \Big)  
\\
& \quad  + h_{W\! B} 
\frac{c_1}{s_1} \big( g^{\mu_1 \mu_2} q_1^{\mu_3}-g^{\mu_1 \mu_3} q_1^{\mu_2}
\big)  \\
& \quad  + h_{\tilde{W}\! B} 
 \frac{c_1}{s_1} \epsilon^{\mu_1 \mu_2 \mu_3 \sigma} q_{1 \,
  \sigma}\,,
\end{split}
\\[1ex]
\begin{split}
\Gamma_{\rm an}^{(4) \, \mu_1 \mu_2 \mu_3 \mu_4} & = 
- 6 h_W\,\sqrt{2}G_\text{F}\,\frac{s_1}{e}\bigg(  
g^{\mu_1 \mu_2} g^{\mu_3 \mu_4} (q_1+q_2)\!\cdot\!(q_3+q_4)
\\ 
& 
\hphantom{- 6 h_W\,\sqrt{2}G_\text{F}\,\frac{s_1}{e}\bigg(} \quad 
 - g^{\mu_2 \mu_3}
g^{\mu_1 \mu_4} 
(q_2\!\cdot\! q_4 + q_1\!\cdot\! q_3)  
\\
& 
\hphantom{- 6 h_W\,\sqrt{2}G_\text{F}\,\frac{s_1}{e}\bigg(} \quad 
  - g^{\mu_1 \mu_3} g^{\mu_2 \mu_4} (q_1\!\cdot\! q_4 + q_2\!\cdot\! q_3)
 \\
& \hphantom{- 6 h_W\,\sqrt{2}G_\text{F}\,\frac{s_1}{e}\bigg(} \quad 
+  g^{\mu_1 \mu_3}  \Big( q_1^{\mu_4} (q_4-q_3)^{\mu_2}
+  q_2^{\mu_4}  q_3^{\mu_2} +  q_1^{\mu_2}  q_3^{\mu_4} \Big)  \\
& \hphantom{- 6 h_W\,\sqrt{2}G_\text{F}\,\frac{s_1}{e}\bigg(} \quad 
 + g^{\mu_1 \mu_4}  \Big( q_1^{\mu_3} (q_3-q_4)^{\mu_2}
+  q_2^{\mu_3}  q_4^{\mu_2} +  q_1^{\mu_2}  q_4^{\mu_3} \Big)  \\
& \hphantom{- 6 h_W\,\sqrt{2}G_\text{F}\,\frac{s_1}{e}\bigg(} \quad 
 + g^{\mu_2 \mu_3}  \Big( q_2^{\mu_4} (q_4-q_3)^{\mu_1}
+  q_1^{\mu_4}  q_3^{\mu_1} +  q_2^{\mu_1}  q_3^{\mu_4} \Big)  \\
& \hphantom{- 6 h_W\,\sqrt{2}G_\text{F}\,\frac{s_1}{e}\bigg(} \quad 
 + g^{\mu_2 \mu_4}  \Big( q_2^{\mu_3} (q_3-q_4)^{\mu_1}
+  q_1^{\mu_3}  q_4^{\mu_1} +  q_2^{\mu_1}  q_4^{\mu_3} \Big)  \\
& \hphantom{- 6 h_W\,\sqrt{2}G_\text{F}\,\frac{s_1}{e}\bigg(} \quad 
  -  g^{\mu_1 \mu_2} \Big( (q_1+q_2)^{\mu_3} q_3^{\mu_4} 
+ (q_1+q_2)^{\mu_4} q_4^{\mu_3} \Big)   \\
& \hphantom{- 6 h_W\,\sqrt{2}G_\text{F}\,\frac{s_1}{e}\bigg(} \quad 
  -  g^{\mu_3 \mu_4} \Big( (q_3+q_4)^{\mu_2} q_2^{\mu_1} 
+ (q_3+q_4)^{\mu_1} q_1^{\mu_2} \Big) \bigg)  \\
& \quad  - 6 h_{\tilde{W}}\,\sqrt{2}G_\text{F}\, \frac{s_1}{e} \Big( g^{\mu_1 \mu_3} 
\epsilon^{\mu_2 \mu_4 \rho \sigma} q_{2 \, \rho} q_{3 \, \sigma} + g^{\mu_1
  \mu_4} \epsilon^{\mu_2 \mu_3 \rho \sigma} q_{2 \, \rho} q_{4 \,
\sigma}
 \\
&  \;\hphantom{- 6 h_{\tilde{W}}\,\sqrt{2}G_\text{F}\, \frac{s_1}{e} \Big(}
 +  (q_4-q_3)^{\mu_1} \epsilon^{\mu_2 \mu_3 \mu_4 \sigma}
q_{2 \, \sigma} + g^{\mu_2 \mu_3}  \epsilon^{\mu_1 \mu_4 \rho \sigma} q_{1 \,
  \rho} q_{3 \, \sigma}  
\\
& \;\hphantom{- 6 h_{\tilde{W}}\,\sqrt{2}G_\text{F}\, \frac{s_1}{e} \Big(}
+ (q_4-q_3)^{\mu_2} \epsilon^{\mu_1 \mu_3 \mu_4 \sigma}
q_{1 \, \sigma} + g^{\mu_2 \mu_4}  \epsilon^{\mu_1 \mu_3 \rho \sigma} 
q_{1 \, \rho} q_{4 \, \sigma}  
\\
& \; \hphantom{- 6 h_{\tilde{W}}\,\sqrt{2}G_\text{F}\, \frac{s_1}{e} \Big(}
+  g^{\mu_3 \mu_4} \epsilon^{\mu_1 \mu_2 \rho \sigma}
(q_2-q_1)_{\rho} (q_3+q_4)_{\sigma}  
\\
& \; \hphantom{- 6 h_{\tilde{W}}\,\sqrt{2}G_\text{F}\, \frac{s_1}{e} \Big(}
+ q_4^{\mu_3} \epsilon^{\mu_1 \mu_2 \mu_4 \sigma} 
(q_2-q_1)_{\sigma} + q_3^{\mu_4} \epsilon^{\mu_1 \mu_2 \mu_3 \sigma} 
(q_2-q_1)_{\sigma} \Big)\,, 
\end{split}
\\[1ex]
\begin{split}
\Gamma_{\rm an}^{(\gamma \gamma H) \, \mu_1 \mu_2} & =  2 \big(h_{\varphi W\! B} - 2 h_{W\! B} s_1
  c_1 \big) \, \big(q_1^{\mu_2} q_2^{\mu_1}-g^{\mu_1 \mu_2} q_1\!\cdot\! q_2\big) \\
& \quad  + 4 \big( h_{\varphi \tilde{W}\!\tilde{B}} - h_{\tilde{W}
\mskip -3mu B} s_1 c_1 \big) \;\epsilon^{\mu_1\mu_2\rho\sigma} q_{1
\,\rho} q_{2 \,\sigma}\,. 
\end{split}
\end{align}
Here the centred dots denote invariant four products.  The Feynman rules
for the $W$-boson decays in the $P_W$~scheme are identical to the SM~ones.


\section{Particle momenta and polarisations}
\label{sec-pola}
\setcounter{equation}{0}

The production process \mbox{$\gamma \gamma \rightarrow WW$} is calculated in
the c.m.~system of the two photons.  Our coordinates are defined in
Sect.~\ref{sec-cross}.  For the incoming photons and outgoing $W$~bosons we
use the polarisation vectors
\begin{align}
\epsilon^{\mu}_1(\lambda_1) & =  \frac{1}{\sqrt{2}} (0,-\lambda_1,-i,0)\,, \\
\epsilon^{\mu}_2(\lambda_2) & =  \frac{1}{\sqrt{2}} (0,\lambda_2,-i,0)\,, \\
\epsilon^{\mu}_3(\lambda_3)^* & =  \frac{1}{\sqrt{2}} 
(0,-\lambda_3 \cos \Theta,i,\lambda_3 \sin \Theta)\,, \\
\epsilon^{\mu}_4(\lambda_4)^* & =  \frac{1}{\sqrt{2}} 
(0,\lambda_4 \cos \Theta,i,-\lambda_4 \sin \Theta)\,, \\
\epsilon^{\mu}_3(0)^* & =  \frac{\sqrt{s}}{2 m_W}
(\beta,\sin \Theta,0,\cos \Theta)\,,\\
\epsilon^{\mu}_4(0)^* & =  \frac{\sqrt{s}}{2 m_W}
(\beta,-\sin \Theta,0,-\cos \Theta)\,,
\end{align}
where $\beta = (1-4m_W^2/s)^{1/2}$ and the asterisk denotes complex
conjugation.  The helicities $\lambda_1$ and $\lambda_2$ of the
photons and $\lambda_3$ and $\lambda_4$ of the $W$~bosons take {\em
only} the values $+1$ or $-1$ here.  The vectors of longitudinal
polarisation of the $W$s are separately given in the last two
equations.  In the frames specified in Sect.~\ref{sec-cross} (see
(\ref{eq-diffs}) and the following equations), the four-momenta of the
final-state fermions are
\begin{align}
p_1^{\mu}&=\frac{1}{2}\sqrt{k_3^2}\;(1,\sin \vartheta \cos \varphi,
\sin \vartheta \sin \varphi,\cos \vartheta)\,,
\\[.2cm]
p_2^{\mu}&=\frac{1}{2}\sqrt{k_3^2}\;(1,-\sin \vartheta \cos \varphi,
-\sin \vartheta \sin \varphi,-\cos \vartheta)\,,
\\[.2cm]
p_3^{\mu}&=\frac{1}{2}\sqrt{k_4^2}\;(1,-\sin \overline{\vartheta}
\cos \overline{\varphi}, -\sin \overline{\vartheta} \sin
\overline{\varphi},-\cos \overline{\vartheta})\,,
\\[.2cm]
p_4^{\mu}&=\frac{1}{2}\sqrt{k_4^2}\;(1,\sin \overline{\vartheta} \cos
\overline{\varphi},
\sin \overline{\vartheta} \sin \overline{\varphi},\cos \overline{\vartheta})\,.
\end{align}
Here we neglected all masses of the final state fermions.


\section{Helicity amplitudes}
\label{sec-helggww}
\setcounter{equation}{0}

In this section we list the helicity amplitudes for \mbox{$\gamma \gamma
\rightarrow WW$} in~(\ref{eq-prodpro}) and the relevant \mbox{$l$-functions}
describing the $W$-decay amplitudes in~(\ref{eq-decay12})
and~(\ref{eq-decay34}).

In the helicity amplitudes \mbox{$\mathcal{M}(\lambda_1, \lambda_2; \lambda_3,
\lambda_4)$} in~(\ref{eq-prodpro}) altogether 36~combinations of the photon
helicities $\lambda_1$ and $\lambda_2$ and $W$-boson helicities $\lambda_3$
and $\lambda_4$ are possible.  In order to give the amplitudes in a convenient
way we distinguish between longitudinal and transverse $W$~helicities.  Thus,
in the following
\be
\lambda_1, \lambda_2, \lambda_3, \lambda_4 = \pm 1\,,
\ee
whereas the amplitudes where one or both $W$~bosons have longitudinal
polarisation are listed separately.  The amplitudes
\mbox{$\mathcal{M}(\lambda_1, \lambda_2; 0, \lambda_4)$} can be obtained from
the amplitudes $\mathcal{M}(\lambda_1, \lambda_2; \lambda_3,0)$ by the
replacements \mbox{$(\lambda_3 \rightarrow \lambda_4)$} and \mbox{$(\lambda_1
\leftrightarrow \lambda_2)$}.  We use the abbreviations
\begin{align}
A &=\left(1 - \beta^2 \cos^2 \Theta \right)\,,
&B&= s-m_H^2\,,
\\
\beta&=\sqrt{1-4\,m_W^2/s}\,,
&\gamma&=\frac{1}{\sqrt{1-\beta^2}}=\frac{\sqrt{s}}{2\,m_W}\,.
\end{align}
For the SM~amplitudes we obtain 
\begin{align}
\mathcal{M}_{\rm SM}(\lambda_1,\lambda_2;0,0) & = \frac{i e^2}{\gamma^2 A} 
\big((\gamma^2+1)(1-\lambda_1\lambda_2)
\sin^2 \Theta -(1+\lambda_1\lambda_2)\big)\,,
\\
\mathcal{M}_{\rm SM}(\lambda_1,\lambda_2;\lambda_3,0) & =  
\frac{-i e^2 \sqrt{2}}{\gamma \; A}(\lambda_1-\lambda_2)(1+\lambda_1 
\lambda_3 \cos \Theta) \sin \Theta\,,
\\
\begin{split}
\mathcal{M}_{\rm SM}(\lambda_1,\lambda_2;\lambda_3,\lambda_4) & = 
\frac{-i e^2}{2 A}
\Big(2 \beta(\lambda_1+\lambda_2)(\lambda_3+\lambda_4)\\
&  \hphantom{=\frac{-i e^2}{2 A}\Big(}
- \gamma^{-2}(1+\lambda_3 \lambda_4)\big(2 \lambda_1 
\lambda_2 + (1-\lambda_1 \lambda_2)\cos^2 \Theta\big) \\
&  \hphantom{=\frac{-i e^2}{2 A}\Big(}  
+ (1+\lambda_1 \lambda_2 \lambda_3 \lambda_4)
(3+\lambda_1 \lambda_2)\\
&  \hphantom{=\frac{-i e^2}{2 A}\Big(}   
+ 2 (\lambda_1-\lambda_2)(\lambda_3-\lambda_4)
\cos \Theta \\
&  \hphantom{=\frac{-i e^2}{2 A}\Big(}  
+ (1-\lambda_1 \lambda_2)(1-\lambda_3
\lambda_4)
\cos^2 \Theta \Big)\,.
\end{split}
\end{align}
These results agree with \cite{Baillargeon:1997rz} apart from an
overall factor of~\mbox{$-i$}. Note their different definition
of~$\gamma$ (they use $\gamma=s/m_W^2$).

\clearpage
The following results are obtained for the parts of the amplitudes
(\ref{eq-Mdecomp}) that multiply an anomalous coupling where the
subscript denotes the corresponding coupling, see (\ref{s1c1}) for
the coefficients $s_1$ and $c_1$:
\begin{align}
\mathcal{M}_{W}(\lambda_1,\lambda_2;0,0) & =  \frac{3 i e s s_1 \sqrt{2}
  G_{\rm F}}{\gamma^2 A} \sin^2 \Theta(1+\lambda_1 \lambda_2)\,,
\\[1ex]
\begin{split}
\mathcal{M}_{W}(\lambda_1,\lambda_2;\lambda_3,0) & =  \frac{3 i e s s_1
  G_{\rm F}}{2 \gamma A} \sin \Theta \\
&\hspace{-1cm} \times \big(
(\lambda_1-\lambda_2)\beta^2 - \beta \cos \Theta (\lambda_1+\lambda_2)
- 2 \lambda_3 \cos \Theta (\lambda_1 \lambda_2 +
\gamma^{-2}) \big)\,,
\end{split}
\\[1ex]
\begin{split}
\mathcal{M}_{W}(\lambda_1,\lambda_2;\lambda_3,\lambda_4) & = 
\frac{3 i e s s_1 \sqrt{2} G_{\rm F}}{4 A}
\Big(-\gamma^{-2}\beta (1+\cos^2 \Theta) (\lambda_1 + \lambda_2) 
(\lambda_3 + \lambda_4)
\\ & \hspace{-1cm} 
+ 2 \sin^2 \Theta \; \big(3+\lambda_3\lambda_4
+\lambda_1 \lambda_2(1-\lambda_3\lambda_4)-\beta (\lambda_1 + \lambda_2)
(\lambda_3 + \lambda_4)\big)  
\\ & \hspace{-1cm}
 -  2 \gamma^{-2}\big( 2+(1-\lambda_1\lambda_2)
\lambda_3 \lambda_4 -\cos^2 \Theta (3+\lambda_1\lambda_2+2\lambda_3\lambda_4) 
\big)\Big)\,, 
\end{split}
\\[2ex]
\mathcal{M}_{\tilde{W}}(\lambda_1,\lambda_2;0,0) & =  
-\frac{3 e s s_1 \sqrt{2} G_{\rm F}}{\gamma^2 A}\sin^2 \Theta (\lambda_1+\lambda_2)\,,
\\[1ex]
\begin{split}
\mathcal{M}_{\tilde{W}}(\lambda_1,\lambda_2;\lambda_3,0) & =\frac{3 e s s_1
  G_{\rm F}}{2 \gamma A} \sin \Theta 
\\ & \hspace{-1cm} \times \Big( \beta(\lambda_1-\lambda_2)\lambda_3 + \cos
  \Theta \; \big(2 \beta + (2 - \beta^2)(\lambda_1+\lambda_2)\lambda_3\big)
  \Big)\,, 
\end{split}
\\[1ex]
\begin{split}
\mathcal{M}_{\tilde{W}}(\lambda_1,\lambda_2;\lambda_3,\lambda_4) & = 
-\frac{3 e s s_1 G_{\rm F}}{\sqrt{2} A} 
\bigg( 2 \sin^2 \Theta (\lambda_1 + \lambda_2 - \beta (\lambda_3 + \lambda_4))
\\ & \hspace{-1cm}
+ \gamma^{-2} \Big( 
(\lambda_1 + \lambda_2) \big(\cos^2 \Theta (2 + \lambda_3 \lambda_4) - 1\big)
\\ & \hspace{-1cm} \hphantom{+\gamma^{-2}\Big(}
- \beta (\cos^2 \Theta + \lambda_1 \lambda_2)
(\lambda_3 + \lambda_4) \Big) \bigg)\,, 
\end{split}
\\[2ex]
\mathcal{M}_{\varphi W}(\lambda_1,\lambda_2;0,0) & =  -\frac{i s^2 s_1^2
  \sqrt{2} G_{\rm F}}{4 B} (1 + \beta^2) (1 + \lambda_1\lambda_2)\,,
\\[1ex] 
\mathcal{M}_{\varphi W}(\lambda_1,\lambda_2;\lambda_3,0) & =  0\,,
\\[1ex]
\mathcal{M}_{\varphi W}(\lambda_1,\lambda_2;\lambda_3,\lambda_4) & =  
-\frac{i s^2 s_1^2 \sqrt{2} G_{\rm F}}{8 \gamma^2 B} (1 + \lambda_1 \lambda_2) 
(1 + \lambda_3 \lambda_4)\,,
\end{align}

\begin{align}
\mathcal{M}_{\varphi \tilde{W}} & =   2 i \lambda_1\, \mathcal{M}_{\varphi
  W}\,,
\\[1ex]
\mathcal{M}_{\varphi B} & =  \frac{c_1^2}{s_1^2}\, \mathcal{M}_{\varphi
  W}\,,
\\[1ex]
\mathcal{M}_{\varphi \tilde{B}}  & =   \frac{c_1^2}{s_1^2}
\,\mathcal{M}_{\varphi \tilde{W}}\,, 
\\[2ex]
\begin{split}
\mathcal{M}_{W\! B}(\lambda_1,\lambda_2;0,0) & =  \frac{2 i
e^2}{A} \frac{c_1}{s_1}
\big( 1 - \lambda_1 \lambda_2 -2 \cos^2 \Theta
 -\gamma^2 (1 + \lambda_1 \lambda_2) \sin^2 \Theta \big)
\\ 
&\quad + \frac{i s^2 G_{\rm F}}{\sqrt{2} B} s_1 c_1 (1 + \beta^2)
(1 + \lambda_1 \lambda_2)\,, 
\end{split}
\\[1ex]
\begin{split}
\mathcal{M}_{W\! B}(\lambda_1,\lambda_2;\lambda_3,0) &	  = 
\frac{i e^2 \gamma}{\sqrt{2} A} \frac{c_1}{s_1} \sin \Theta
\Big( (\lambda_2-\lambda_1)(1 + \gamma^{-2})
\\ & 
\quad
+ \big( \beta (\lambda_1 + \lambda_2) +
2 \lambda_3 (\lambda_1 \lambda_2 - \gamma^{-2}) \big) \cos \Theta 
\Big)\,, 
\end{split}
\\[1ex] 
\begin{split}
\mathcal{M}_{W\! B}(\lambda_1,\lambda_2;\lambda_3,\lambda_4) & =  
-\frac{i e^2}{2 A} \frac{c_1}{s_1} \bigg( \beta (\lambda_1
+ \lambda_2) (\lambda_3 + \lambda_4) (1 + \cos^2 \Theta)
\\ &  \hphantom{-\frac{i e^2}{2 A} \frac{c_1}{s_1} \bigg(}
+ 2 \Big( 2 + (\lambda_1-\lambda_2) (\lambda_3-\lambda_4) 
\cos \Theta 
\\ &  \hphantom{-\frac{i e^2}{2 A} \frac{c_1}{s_1} \bigg(}
 + \big( (\lambda_1\lambda_2 - 1) \cos^2 \Theta + 1 +
\lambda_1 \lambda_2 \big) \lambda_3 \lambda_4 \Big) \bigg)  
\\ &\quad + \frac{i s^2 \sqrt{2} G_{\rm F}}{4 \gamma^2 B} s_1 c_1 (1 + \lambda_1 \lambda_2)
(1 + \lambda_3 \lambda_4)\,,  
\end{split}
\\[2ex]
\label{eq-amwtb1}
\mathcal{M}_{\tilde{W}\! B}(\lambda_1,\lambda_2;0,0) & =  
2 e^2 \frac{c_1}{s_1} \gamma^2 (\lambda_1 + \lambda_2) -
\frac{s^2 G_{\rm F}}{\sqrt{2} B} s_1 c_1 (1 + \beta^2) (\lambda_1 +
\lambda_2)\,,
\\[1ex]
\begin{split}
\label{eq-amwtb2}
\mathcal{M}_{\tilde{W}\! B}(\lambda_1,\lambda_2;\lambda_3,0)
& =  \frac{e^2 \gamma}{\sqrt{2} A} \frac{c_1}{s_1} \sin \Theta
\\ & \quad\times\Big( \beta (\lambda_2 - \lambda_1) \lambda_3
- \cos \Theta \big( 2 \beta + \beta^2 (\lambda_1 + \lambda_2) \lambda_3 \big) 
\Big)\,,
\end{split}
\\[1ex]
\begin{split}
\label{eq-amwtb3}
\mathcal{M}_{\tilde{W}\! B}(\lambda_1,\lambda_2;\lambda_3,\lambda_4) 
& =  \frac{e^2}{A} \frac{c_1^2}{s_1} \big( \lambda_3 (\lambda_1 + 
\lambda_2) + \beta (\lambda_1 \lambda_2 + \cos^2 \Theta) \big)(\lambda_3 + \lambda_4) 
\\ &\quad  
- \frac{s^2 \sqrt{2} G_{\rm F}}{4 \gamma^2 B} s_1 c_1^2 (\lambda_1 +
\lambda_2) (1 + \lambda_3 \lambda_4)\,.  
\end{split}
\end{align}
The $l$-functions used in~(\ref{eq-decay12}) and~(\ref{eq-decay34})
are given by
\begin{alignat}{5}
&l_- \; & = & \; d_+(\vartheta) e^{-i\varphi}\,, \hspace{1cm}
&l_0 \; & = & \; - d_0(\vartheta)\,, \hspace{1cm}
&l_+ \; & = & \; d_-(\vartheta)e^{i\varphi}\,,\\
&\overline{l}_- \; & = & \; d_+(\overline{\vartheta}) e^{i\overline{\varphi}}\,, \hspace{1cm} 
&\overline{l}_0 \; & = & \; - d_0(\overline{\vartheta})\,, \hspace{1cm}
&\overline{l}_+ \; & = & \; d_-(\overline{\vartheta})e^{-i\overline{\varphi}}
\end{alignat}
with $d_{\pm}(x)=(1\pm\cos x)/\sqrt{2}$ and $\;d_0(x)=\sin x$.

\end{appendix}



\begin{thebibliography}{99}

\bibitem{Richard:2001qm}
``TESLA Technical Design Report Part I: Executive Summary,''
eds.\ F.~Richard, J.~R.~Schneider, D.~Trines and A.~Wagner, 
DESY, Hamburg, 2001 [hep-ph/0106314];\\
%
``TESLA Technical Design Report Part III: Physics at an $e^+e^-$
Linear Collider,''  eds.\ R.-D.~Heuer, D.~Miller, F.~Richard,
P.~M.~Zerwas, DESY, Hamburg, 2001 [hep-ph/0106315].

\bibitem{Ellis:1998wx}
J.~R.~Ellis, E.~Keil and G.~Rolandi,
``Options for Future Colliders at CERN,''
CERN-EP-98-03;\\
%
J.~P.~Delahaye {\it et al.},
``CLIC---a Two-Beam Multi-TeV $e^+ e^-$ Linear Collider,''
in: {\it Proc.\ of the 20th Intl.\ Linac Conference LINAC 2000 }
ed.\ Alexander W.~Chao, eConf {C000821}, MO201 (2000) [physics/0008064];\\
%
E.~Accomando {\it et al.}  [CLIC Physics Working Group Collaboration],
``Physics at the CLIC multi-TeV linear collider,'' 
CERN, Geneva, 2004, [arXiv:hep-ph/0412251].

\bibitem{Badelek:2001xb}
``TESLA Technical Design Report, Part VI, Chapter 1: Photon Collider at
TESLA,'' B.~Badelek {\it et al.}, DESY, Hamburg, 2001, [hep-ex/0108012].

\bibitem{Burkhardt:2002vh}
H.~Burkhardt and V.~Telnov,
``CLIC 3-TeV Photon Collider Option,''
CERN-SL-2002-013-AP.

\bibitem{Buchmuller:1985jz}
W.~Buchm\"uller and D.~Wyler,
Nucl.\ Phys.\ B {\bf 268}, 621 (1986).

\bibitem{Leung:1984ni}
C.~N.~Leung, S.~T.~Love and S.~Rao,
Z.\ Phys.\ C {\bf 31}, 433 (1986).

\bibitem{Nachtmann:2004ug}
O.~Nachtmann, F.~Nagel and M.~Pospischil,
Eur.\ Phys.\ J.\ C {\bf 42} (2005) 139
[arXiv:hep-ph/0404006].


\bibitem{Hagiwara:1992eh}
K.~Hagiwara, S.~Ishihara, R.~Szalapski and D.~Zeppenfeld,
Phys.\ Lett.\ B {\bf 283}, 353 (1992);
\\
K.~Hagiwara, S.~Ishihara, R.~Szalapski and D.~Zeppenfeld,
Phys.\ Rev.\ D {\bf 48}, 2182 (1993).



\bibitem{Hagiwara:1986vm}
K.~Hagiwara, R.~D.~Peccei, D.~Zeppenfeld and K.~Hikasa,
Nucl.\ Phys.\ B {\bf 282}, 253 (1987).


\bibitem{Diehl:2002nj}
M.~Diehl, O.~Nachtmann and F.~Nagel,
Eur.\ Phys.\ J.\ C {\bf 27}, 375 (2003)
[hep-ph/0209229];\\
%
%
M.~Diehl, O.~Nachtmann and F.~Nagel,
Eur.\ Phys.\ J.\ C {\bf 32}, 17 (2003)
[hep-ph/0306247].

\bibitem{Diehl:1993br}
M.~Diehl and O.~Nachtmann,
Z.\ Phys.\ C {\bf 62}, 397 (1994);\\
%
M.~Diehl and O.~Nachtmann,
Eur.\ Phys.\ J.\ C {\bf 1}, 177 (1998)
[hep-ph/9702208].


\bibitem{Gaemers:1978hg}
K.~J.~F.~Gaemers and G.~J.~Gounaris,
Z.\ Phys.\ C {\bf 1}, 259 (1979).


\bibitem{Berends:1997av}
F.~A.~Berends {\it et al.},
``Report of the working group on the measurement of triple gauge boson
couplings,''
J.\ Phys.\ G {\bf 24}, 405 (1998)
[hep-ph/9709413].


\bibitem{Menges:2001gg}
W.~Menges,
``A study of charged current triple gauge couplings at TESLA,''
LC-PHSM-2001-022.

\bibitem{Abe:2001wn} 
T.~Abe {\it et al.}  [American Linear Collider Working Group Collaboration],
in {\it Proc. of the APS/DPF/DPB Summer Study on the Future of Particle
  Physics (Snowmass 2001)} ed. N.~Graf, SLAC-R-570
{\it Resource book for Snowmass 2001, 30 Jun - 21 Jul 2001, Snowmass, Colorado}.




\bibitem{Kuss:1997mf}
I.~Kuss and E.~Nuss,
Eur.\ Phys.\ J.\ C {\bf 4}, 641 (1998)
[hep-ph/9706406].

\bibitem{Bozovic-Jelisavcic:2002ta}
I.~Bo\v{z}ovi\'c-Jelisav\v{c}i\'c, K.~M\"onig and J.~\v{S}ekari\'c,
 ``Measurement of trilinear gauge couplings at a $\gamma\gamma$ and $e\gamma$
collider,''
hep-ph/0210308;\\
K.~M\"onig, ``Electroweak Gauge Theories and Alternative Theories at a Future
Linear \mbox{$e^+ e^-$}~Collider,''
hep-ph/0309021.


\bibitem{Tupper:1980bw}
G.~Tupper and M.~A.~Samuel,
Phys.\ Rev.\ D {\bf 23} (1981) 1933;\\
%
S.~Y.~Choi and F.~Schrempp,
Phys.\ Lett.\ B {\bf 272} (1991) 149;\\
%
E.~Yehudai,
Phys.\ Rev.\ D {\bf 44} (1991) 3434.

\bibitem{Belanger:1992qi}
G.~B\'elanger and F.~Boudjema,
Phys.\ Lett.\ B {\bf 288} (1992) 210.
 
\bibitem{Marfin:2003jg}
I.~B.~Marfin, V.~A.~Mossolov and T.~V.~Shishkina,
``Anomalous quartic boson couplings via $\gamma\gamma\to W^+ W^-$ and  $\gamma\gamma\to W^+ W^- Z$ at the TESLA kinematics,''
hep-ph/0304250.
%
\bibitem{Belanger:hp}
G.~B\'elanger and G.~Couture,
Phys.\ Rev.\ D {\bf 49} (1994) 5720.
%

\bibitem{Baillargeon:1997rz}
M.~Baillargeon, G.~B\'elanger and F.~Boudjema,
Nucl.\ Phys.\ B {\bf 500} (1997) 224
[hep-ph/9701372].

\bibitem{Bredenstein:2004ef}
A.~Bredenstein, S.~Dittmaier and M.~Roth,
Eur.\ Phys.\ J.\ C {\bf 36} (2004) 341
[hep-ph/0405169].

\bibitem{Bredenstein:2005zk}
A.~Bredenstein, S.~Dittmaier and M.~Roth,
Eur.\ Phys.\ J.\ C {\bf 44} (2005) 27
hep-ph/0506005.

\bibitem{Banin:1998ap}
A.~T.~Banin, I.~F.~Ginzburg and I.~P.~Ivanov,
Phys.\ Rev.\ D {\bf 59} (1999) 115001
[arXiv:hep-ph/9806515]; \\
E.~Gabrielli, V.~A.~Ilyin and B.~Mele,
Phys.\ Rev.\ D {\bf 60} (1999) 113005
[arXiv:hep-ph/9902362].

\bibitem{Choi:1996xt}
S.~Y.~Choi, K.~Hagiwara and M.~S.~Baek,
Phys.\ Rev.\ D {\bf 54} (1996) 6703
[hep-ph/9605334].

\bibitem{Gounaris:1995mc}
G.~J.~Gounaris, J.~Layssac and F.~M.~Renard,
Z.\ Phys.\ C {\bf 69} (1996) 505
[hep-ph/9505430].

\bibitem{Nachtmann:opti}
O.~Nachtmann, F.~Nagel, M.~Pospischil and A.~Utermann, ``Effective-Lagrangian approach 
        to $\gamma\gamma\rightarrow WW$;
II: Results and comparison with $e^+e^- \rightarrow WW$,'' hep-ph/0508133.

\bibitem{Atwood:1991ka}
D.~Atwood and A.~Soni,
Phys.\ Rev.\ D {\bf 45}, 2405 (1992);\\
%
M.~Davier, L.~Duflot, F.~Le Diberder and A.~Roug{\'e},
Phys.\ Lett.\ B {\bf 306}, 411 (1993).


\bibitem{Nachtmann:2004fy}
O.~Nachtmann and F.~Nagel,
Eur.\ Phys.\ J.\ C {\bf 40} (2005) 497
[hep-ph/0407224].	

\bibitem{Nachtmann:ta}
O.~Nachtmann,
``Elementary Particle Physics: Concepts and Phenomena,'' 
Springer Verlag, Berlin, Germany (1990).



\end{thebibliography}
\end{document}